\documentclass[sigconf]{acmart}

\AtBeginDocument{%
  \providecommand\BibTeX{{%
    \normalfont B\kern-0.5em{\scshape i\kern-0.25em b}\kern-0.8em\TeX}}}

\usepackage{color} 
\usepackage{xspace} 
\usepackage{subfigure}
\usepackage{amsfonts}
\usepackage{booktabs}
\usepackage{multirow}
\usepackage{enumitem}
\usepackage{bm}
\usepackage{amsmath}
\usepackage{natbib}
\usepackage{tikz} 
\usepackage{ulem}
\usepackage{color}
\usepackage{graphicx}

\usepackage[ruled,linesnumbered]{algorithm2e}

\newcommand{\zhou}[1]{{\color{blue}#1}}
\newcommand{\zhoucom}[1]{{\color{red}(zhoucom:#1)}} 
\newcommand{\li}[1]{{\color{green} \sout{#1}}} 
\newcommand{\old}[1]{{\color{black}#1}} 
\newcommand{\licom}[1]{{\color{black}#1}} 
\newcommand{\hide}[1]{}

\newcommand{\B}[1]{{\bfseries #1}}
\newcommand{\model}{\textsf{SIGN}\xspace}
\newcommand{\gnn}{PGAL\xspace}
\newcommand{\pool}{PiPool\xspace}

\newcommand{\graph}{\ensuremath{\mathcal{G}_I}}
\newcommand{\proteinV}{\ensuremath{\mathcal{V}^P}}
\newcommand{\ligandV}{\ensuremath{\mathcal{V}^L}}
\newcommand{\proteinM}{\ensuremath{M^P}}
\newcommand{\ligandM}{\ensuremath{M^L}}

\newcommand{\angleD}{\ensuremath{\bm{D_A}}}
\newcommand{\cat}{\ensuremath{\mathbin\Vert}}

\newcommand{\dta}{protein-ligand binding affinity}

\newcommand{\mycaption}[1]{\caption{\normalfont{#1}}}

\copyrightyear{2021} 
\acmYear{2021} 
\setcopyright{acmcopyright}\acmConference[KDD '21]{Proceedings of the 27th ACM SIGKDD Conference on Knowledge Discovery and Data Mining}{August 14--18, 2021}{Virtual Event, Singapore}
\acmBooktitle{Proceedings of the 27th ACM SIGKDD Conference on Knowledge Discovery and Data Mining (KDD '21), August 14--18, 2021, Virtual Event, Singapore}
\acmPrice{15.00}
\acmDOI{10.1145/3447548.3467311}
\acmISBN{978-1-4503-8332-5/21/08}

\begin{document}
\fancyhead{}

\title{Structure-aware Interactive Graph Neural Networks \\ for the Prediction of Protein-Ligand Binding Affinity}

\author{Shuangli Li{$\scriptstyle ^{1,2\dagger}$}, Jingbo Zhou{$\scriptstyle ^{2*}$}, Tong Xu{$\scriptstyle ^{1}$}, Liang Huang{$\scriptstyle ^{4,5}$}, Fan Wang{$\scriptstyle ^{3}$}}
\author{Haoyi Xiong{$\scriptstyle ^{3}$}, Weili Huang{$\scriptstyle ^{3,6}$}, Dejing Dou{$\scriptstyle ^{2*}$}, Hui Xiong{$\scriptstyle^{7*}$}}
\thanks{$\dagger$This work was done when the first author was an intern in Baidu Research under the supervision of the second author.}
\thanks{$^*$Corresponding authors.}

\affiliation{$^{1}$University of Science and Technology of China,$^{2}$Business Intelligence Lab, Baidu Research\country{}}
\affiliation{ $^{3}$Baidu Inc., $^{4}$Baidu Research USA, $^{5}$Oregon State University, $^6$HWL Consulting LLC, $^{7}$Rutgers University\country{}}
\email{lsl1997@mail.ustc.edu.cn, {zhoujingbo, wangfan04, xionghaoyi, doudejing}@baidu.com}
\email{tongxu@ustc.edu.cn, {liang.huang.sh, lwlily99}@gmail.com, hxiong@rutgers.edu}



\begin{abstract} \label{sec-abstract}


Drug discovery often relies on the successful prediction of protein-ligand binding affinity. Recent advances 
have shown great promise in applying graph neural networks (GNNs) for better affinity prediction by learning the representations of protein-ligand complexes. However, existing solutions usually treat protein-ligand complexes as topological graph data, thus\hide{and} the biomolecular structural information\hide{ of atoms} is not fully utilized. 
The essential long-range interactions among atoms are also neglected in\hide{ existing} GNN models. To this end, 
we propose a structure-aware interactive graph neural network (\model) which consists of two components: polar-inspired graph attention layers (\gnn) and pairwise interactive pooling (\pool).  Specifically, \gnn iteratively performs the node-edge aggregation process to update\hide{ the} embeddings of nodes and edges while preserving the distance and angle information among atoms\hide{ in protein-ligand complexes}. Then, \pool is adopted to gather interactive edges\hide{ based on atomic types} with a subsequent reconstruction loss to reflect the global interactions\hide{ in the complex}. 
Exhaustive experimental study on two benchmarks verifies the superiority of \model.



\end{abstract}



\maketitle

\section{Introduction} \label{sec-introduction}

\begin{figure}[t]
\centering
\includegraphics[width=0.9\columnwidth]{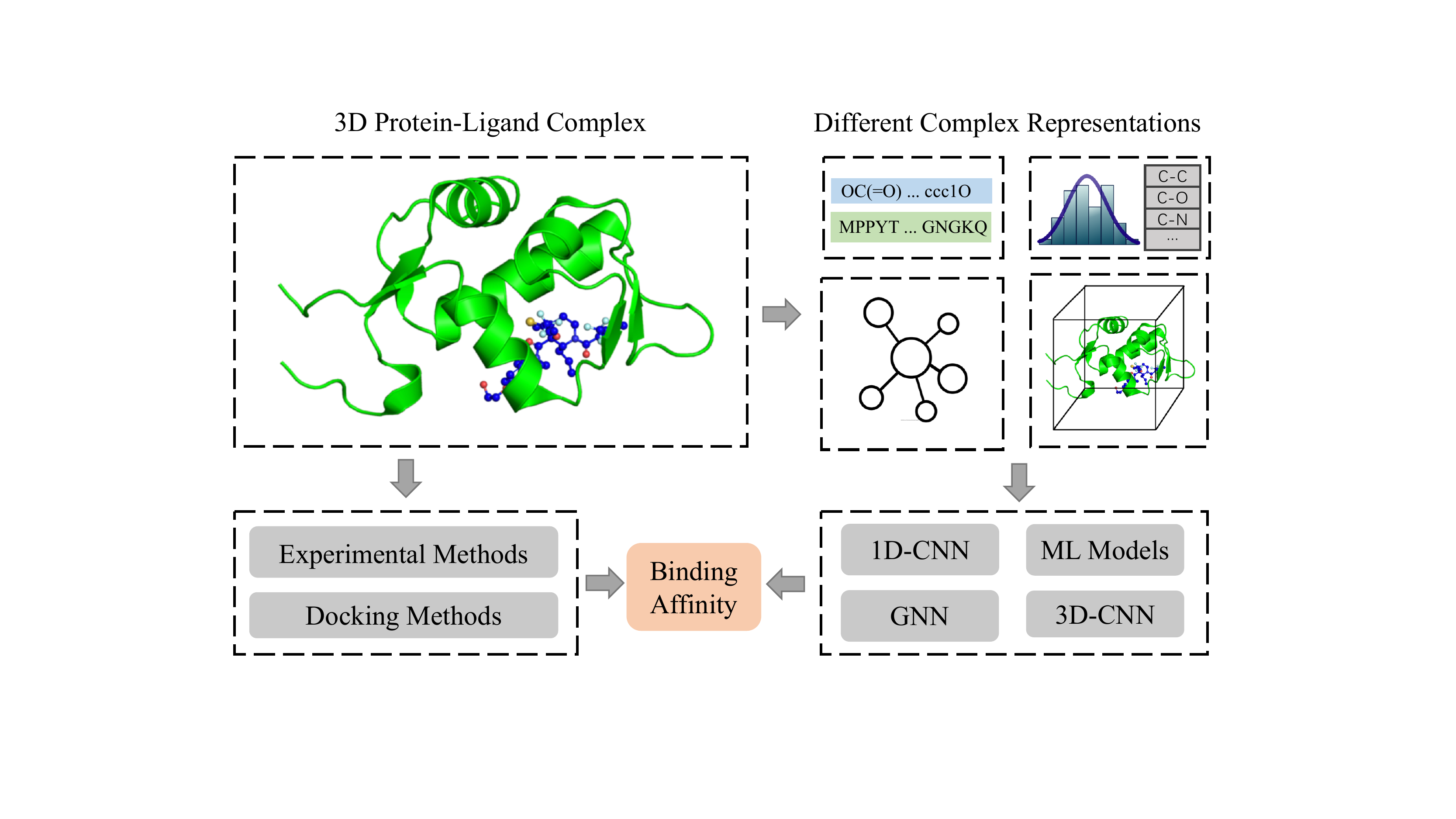}
\vspace{-4mm}
\caption{A brief summary for protein-ligand binding affinity prediction. (1) Top left: An example of protein-ligand complex (PDB: 5HMI). (2) Top right: Various representations of complex. (3) Bottom left: Traditional Methods. (4) Bottom right: Machine learning and deep learning methods.}
\label{fig-intro}
\vspace{-5mm}
\end{figure}

The prediction of protein-ligand binding affinity\hide{Protein-ligand binding affinity prediction} has been widely considered as one of the most important tasks in computational drug discovery \cite{kitchen2004docking}. Here ligands are usually drug candidates including small molecules and biologics which can interact with proteins as agonists\hide{, antagonists} or inhibitors in the biological processes to cure diseases. The binding affinity, defined as the strength of the binding interaction between a protein and a ligand (e.g., drug), can be measured by experimental methods. However, those biological tests are laborious and time-consuming. \hide{Since there are possibly $10^{23}$ potential candidates for developing new drugs \cite{polishchuk2013estimation}, }With the computer aided simulation \hide{models}\old{methods and the data-driven learning models}, binding affinities can be predicted in the early stage of drug discovery. Instead of applying costly biological methods directly to screen numerous candidate molecules, the prediction of binding affinity can help to rank\hide{all } drug candidates and prioritize the appropriate ones for subsequent testing to accelerate the process of drug screening \cite{PMID:28150235}.

With the development of structural biology and protein structure prediction, especially the recent Alphafold II model \cite{callaway2020will}, 
there are growing three-dimensional (3D) structure protein data, which enables a new paradigm for structure-based drug discovery \licom{\cite{jhoti2007structure,meng2011molecular,batool2019structure}}. It has been demonstrated that 3D structural information can effectively contribute to the drug design \cite{leach2006prediction}. Indeed, since there are already many accurate and robust algorithms to find poses of protein-ligand complexes (e.g., binding site prediction methods and docking methods), it is significant to focus on the much harder task of binding affinity prediction\hide{binding affinity prediction task} \cite{ballester2010machine}. \hide{Therefore, }To \hide{ take advantage of different representations}learn\hide{ such} useful 3D structure from a protein-ligand complex, as illustrated in Figure \ref{fig-intro}, many efforts have been devoted to estimating more accurate binding affinity for effective drug design. Docking methods \hide{ruiz2014rdock(hidden)}\cite{jain2003surflex,trott2010autodock,allen2015dock\hide{,zhang2020edock}} play an important role to predict how a specific ligand binds to the target protein with affordable computational costs. While the docking process can identify the binding \hide{site}pose of the protein-ligand complex with relatively high accuracy, its prediction of binding affinity is inaccurate and unreliable \hide{jain2008bias(hidden)}\cite{moitessier2008towards,ballester2010machine} due to poor scoring functions, which limits the applicability of docking methods in drug discovery. 
Compared to docking calculations, traditional machine learning methods \cite{ballester2010machine,kinnings2011machine} have improved the performance by learning the extracted features from protein-ligand complexes. However, these approaches with limited generalizability require expert knowledge and heavily rely on feature engineering.

Recently, deep learning for binding affinity prediction have become\hide{becomes} an emerging research area, which represents the complex as sequence data \cite{ozturk2018deepdta}, 3D grid-like data \cite{wallach2015atomnet} or graph data \cite{10.1093/bioinformatics/btaa921} to employ various neural networks. One of the key challenges of deep learning in structural biology is how to model the 3D spatial structure for better performance. To this end, most of the existing works \hide{wallach2015atomnet}\licom{\cite{wallach2015atomnet,ragoza2017protein, stepniewska2018development}} attempt to apply 3D convolutional neural networks (3D CNNs) by treating the complex as a 3D-grid representation. However, the cost of these models is huge, especially when considering long-range interactions. What's more, both the absence of topological information and the sensitivity to rotation in the complex have a negative effect on the prediction results. 

Despite the powerful ability of graph neural networks (GNNs) to learn graph representations \cite{li2020competitive,zheng2021drug,liu2021vldb}, there are only a few studies \cite{lim2019predicting,10.1093/bioinformatics/btaa921} on\hide{making use of} using GNNs to predict the protein-ligand binding affinity\hide{interactions between protein and ligand}. By contrast, many researchers have greatly developed GNN models in other fields of drug discovery \licom{\cite{sun2020graph,zang2020moflow}}, such as predicting molecular property \cite{yang2019analyzing,maziarka2020molecule,klicpera_dimenet_2020} and chemical reaction \cite{do2019graph}. Nevertheless, these domain-specific models \hide{always}tend to lose their effectiveness when modeling the larger biomolecules, e.g., protein-ligand complexes. In general, most of the existing GNNs in drug design aim to learn the spatial structure by incorporating the distance information, which is insufficient to model the 3D structure of complex. Moreover, the fundamental long-range interactive information between proteins and ligands\hide{ (e.g., Carbon-Carbon co-occurrence interaction)}, which is valuable for predicting the binding affinity \cite{leckband1992long}, cannot be handled under the current GNN framework.

To overcome the above limitations, we propose a novel \underline{S}tructure-aware \underline{I}nteractive \underline{G}raph Neural \underline{N}etwork (\model) to learn the constructed complex graph for predicting the protein-ligand binding affinity. \model is equipped with two designed components to correspondingly address the challenges, namely the \textit{polar-inspired graph attention layers (\gnn)} for modeling 3D spatial structure and the \textit{pairwise interactive pooling (\pool)} for leveraging long-range interactions.
Firstly, the key idea of \gnn is to establish a polar coordinate system for each central target and to preserve both distance and angle information of neighbors when performing the aggregation process. More specifically, we apply the node-edge interactive scheme \hide{recursively}iteratively with graph attention to integrate spatial factors for effectively learning the 3D structure of complex. 

\hide{Considering the large size of protein and the redundancy problem,}\hide{Based on}In view of the large size of the protein, it is \hide{needless}redundant to contain the complete protein structure in the complex graph, but in this way the long-range interactive information between the protein and the ligand is also lost. To deal with this issue, \pool, the secondary part of \model
is designed to incorporate such global interactions into our model, which employs an atomic type-aware pooling process on edges with introducing an auxiliary learning task to reconstruct the atomic interaction matrix. By this means, \model can enhance the representation learning for complexes with involving both 3D spatial structures and global interactions. To summarize, the main contributions of this paper are as follows:
\begin{itemize}[leftmargin=*,topsep=3pt]
    \item To the best of our knowledge, we are among the first to develop graph neural networks from the perspective of polar coordinates for structure-based binding affinity prediction\hide{ problem}. 
    \item We propose a novel structure-aware interactive graph neural network (\model), which can capture not only 3D spatial information through polar-inspired graph attention layers (\gnn), but also global long-range interactions through pairwise interactive pooling (\pool) in a semi-supervised manner.
     \item We conduct extensive experiments using two benchmark datasets to evaluate the performance of the proposed model, which demonstrates the effectiveness of our \model with better generalizability.
\end{itemize}

\vspace{-5mm}

\hide{
\zhou{Protein-ligand binding affinity prediction has been widely considered as one of the most important tasks in computational drug discovery for a long time \cite{drews2000drug}. Here ligands usually are drugs and small molecules which can react with protein.
The binding affinity, which is the quantity of binding strength measured by a real number between protein and ligand, indicates the probability to activate or inhibit a biological process to cure a disease. The prediction of binding affinity is one of the crucial steps in the early stage \cite{kitchen2004docking} of drug discovery.
For example, predicting the binding affinity can help to rank \hide{all }candidate drugs and prioritize the appropriate ones for subsequent testing to accelerate the process of virtual screening \cite{PMID:28150235}. 
How to accurately predict the protein-ligand binding affinity has attracted tenuous research interest in the past decades from both machine learning and computational biology communities \cite{bohm1994development,gohlke2000knowledge, wang2002further,wang2003comparative,sousa2006protein,jacob2008protein,trott2010autodock,colwell2018statistical,ozturk2018deepdta,ozturk2019widedta}. 
}
With the development of structural biology and protein structure prediction, especially the recent Alphafold II model \cite{callaway2020will} designed by DeepMind group \footnote{https://deepmind.com/research/case-studies/alphafold}, there are 
growing three-dimensional (3D) structure protein data, which enables a new paradigm for structure-based drug discovery \cite{jhoti2007structure,meng2011molecular,batool2019structure}. It has been demonstrated that 3D structure information can effectively contribute to the drug design \cite{leach2006prediction}. During this process, the prediction of binding affinity, which indicates the strength of the interaction between protein and ligand (i.e., drug), is one of the crucial steps in the early stage \cite{kitchen2004docking}.
For example, accurately predicting the binding affinity can help to rank \hide{all }candidate drugs and prioritize the appropriate ones for subsequent testing to accelerate the process of virtual screening \cite{PMID:28150235}. 
}
\section{Related Work} \label{sec-related}
In this section, we first review the related literatures about predicting protein-ligand binding affinity and then detail recent advances in graph neural networks for drug discovery.

{\bfseries Protein-Ligand Binding Affinity Prediction.}
As a crucial stage in drug discovery, predicting protein-ligand binding affinity has been intensively studied for a long time \licom{\cite{sousa2006protein,jacob2008protein}}, which is of great importance for efficient and accurate drug screening. The earlier empirical-based methods \hide{bohm1994development,wang2003comparative(hidden)}\cite{gohlke2000knowledge, wang2002further,trott2010autodock} design docking and scoring functions specially to make predictions, while expert domain knowledge is required to encode internal biochemical interactions. Later on, statistical and machine learning-based methods \cite{colwell2018statistical} are developed to predict binding affinity based on data-driven learning, which attempt to extract protein-ligand features and use classic models for regression, such as random forest \cite{ballester2010machine} and SVM \cite{kinnings2011machine}. These approaches are dependent on the quality of hand-crafted features and lack of generality on the larger dataset. Recently, several deep learning-based models \hide{ozturk2019widedta(hidden)}\cite{ozturk2018deepdta} \old{utilize 1D convolutions and pooling to capture potential patterns} from raw sequence information of both ligand and protein. 
However, only using separate character representations fails to achieve desirable performance.

With the recent advances in predicting structures of proteins \cite{callaway2020will} and the increasing availability of 3D-structure protein-ligand data \cite{wang2005pdbbind}, there is another hot research \hide{line}area of studying structure-based approaches, which focus on learning from 3D-structure protein-ligand complexes to predict binding affinity. Some recent works \cite{ragoza2017protein, stepniewska2018development} represent the protein-ligand complex as 3D grid-like data and \old{use 3D convolutions (3D-CNNs) to take advantage of spatially-local correlations. Though these \hide{3D-CNN }approaches can learn spatial information, one limitation is that positions of proteins and ligands in different complexes are changeable, such as different angle rotations, which means the spatial structure of 3D grid-like modeling is inevitably incomplete.} More recently, OnionNet \cite{zheng2019onionnet} employs CNN models to learn the complex representation from the extracted element-specific interaction features between a protein and its ligand. However, all the above models neglect the critical topological structure information of complex. In the work \cite{lim2019predicting}, a protein-ligand complex is represented as a weighted graph with distance information. Then graph attention networks are applied to predicting the interactions. Nevertheless, only distance information between atoms is not adequate to model 3D-structure interactions. In this paper, we also focus on the structure-based prediction of protein-ligand binding affinity with incorporating abundant spatial information.

{\bfseries Graph Neural Networks for Drug Discovery.}
Inspired by the great advantage of graph neural networks (GNNs) in modeling graph data, much attention has been devoted to applying them in computational drug discovery \cite{sun2020graph}, such as the prediction of molecular property \cite{hao2020asgn} and protein interface \cite{liu2020deep}. Treating the molecule as a graph, GNNs can learn the graph-level representation for drug or protein by aggregating structural information. GraphDTA \cite{10.1093/bioinformatics/btaa921} adopts GNN models \cite{kipf2017semi,velivckovic2018graph,xu2018powerful} to learn drug presentation with combining the protein representation from 1D convolutions to predict binding affinity. In attributed molecular graphs, the edges between atoms contain valuable information, such as distance or bond order. To leverage rich attributes in the molecule, edge-oriented message passing neural networks \cite{yang2019analyzing,song2020communicative,zhou2020distance} are proposed to update both node and edge embeddings \hide{ in an interactive manner}. Meanwhile, there are\hide{is} also some efforts to model the 3D-structure of molecule by improving GNNs with spatial information, such as distance \cite{lim2019predicting,maziarka2020molecule}, angle \cite{klicpera_dimenet_2020}, and 3D coordinate \cite{danel2020spatial}. However, these models fail to consider the spatial interactions between proteins and ligands\hide{ligand and protein}. In addition, the function of learning angle information in \cite{klicpera_dimenet_2020} is designed for density functional theory, which is only beneficial for predicting molecular properties \hide{instead of}rather than protein-ligand binding affinity. To overcome these limitations, we propose an interaction-aware GNN framework with integrating both distance and angle factors harmoniously.
\section{Preliminaries}\label{sec-pre}
In this section, we \hide{first} introduce some definitions used in our model and formulate the structure-based prediction problem for protein-ligand binding affinity\hide{prediction problem}. \hide{Then we describe the process of constructing the complex graph.} The frequently used notations in this paper are summarized in Table \ref{table-symbol}.


\begin{table}[t]
	\mycaption{Mathematical notations.
	}
	\vspace{-3ex}
	\label{table-symbol}
	\begin{tabular}{cl}
		\toprule
		Notation	&	Description	\\
		\midrule
		\proteinV,\ligandV	& The atom node sets of protein and ligand \\
		\proteinM,\ligandM	& The 3D position matrices of protein and ligand \\
		\graph	& The complex interaction graph	\\
		$a_i$	& The $i$-th atom node in \graph \\
		$e_{ij}$ & The directed edge from atom $a_i$ to atom $a_j$ \\
		$\mathcal{N}_{e}(a_i)$ & The neighboring edges of atom $a_i$ \\
		$\mathcal{N}_{e}(e_{ij})$ & The neighboring edges of edge $e_{ij}$ \\
		$\bm{a}_{i}$, $\bm{e}_{ij}$ & The embedding vectors of atom $a_i$ and edge $e_{ij}$  \\
		$\bm{d}_{ij}$ & The spatial embedding vector between $a_i$ and $a_j$ \\
        \bottomrule
        
	\end{tabular}
	\vspace{-3ex}
\end{table}

\begin{figure}[t]
\centering
\includegraphics[width=0.9\columnwidth]{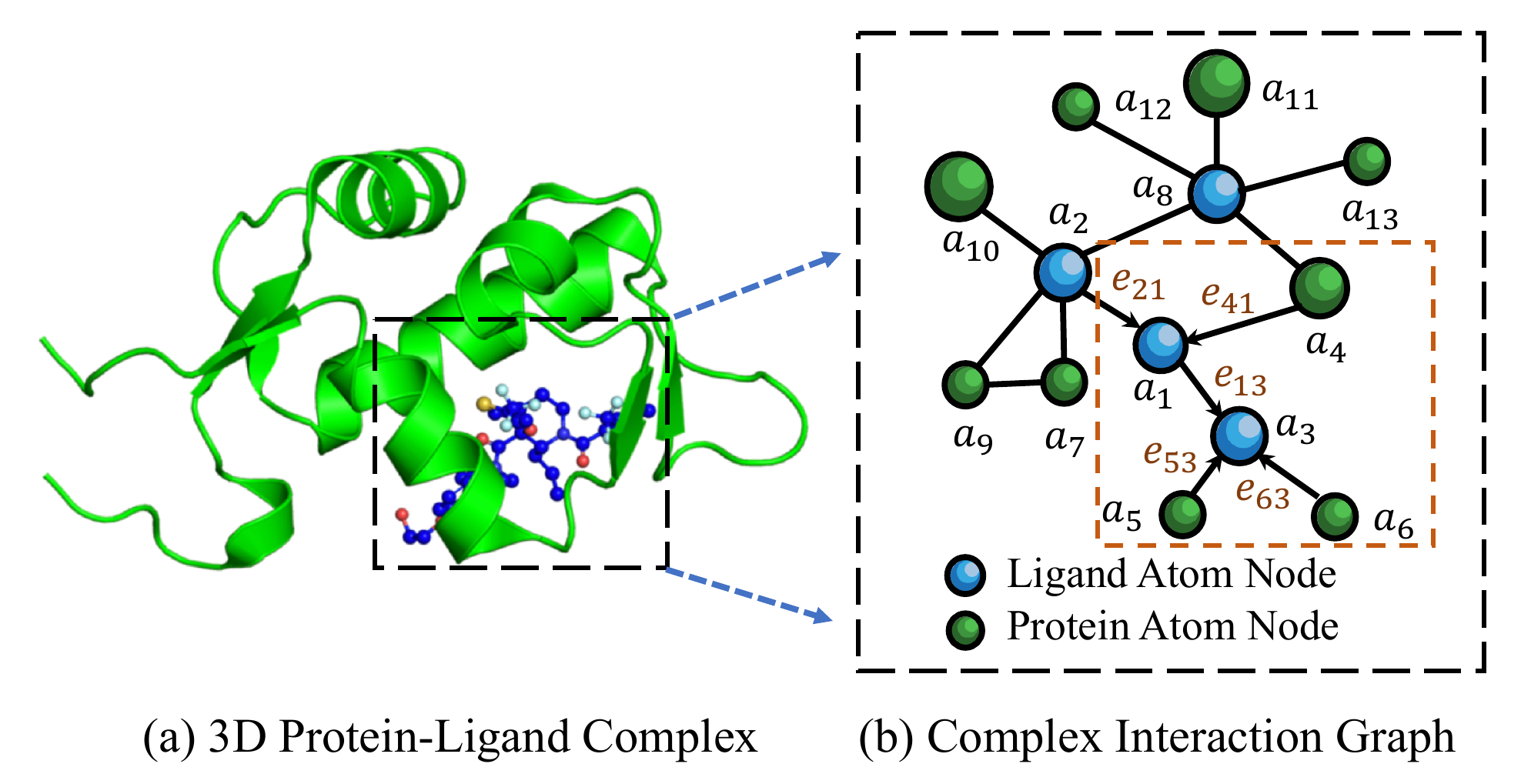}
\vspace{-4mm}
\caption{An illustrative example of converting the protein-ligand complex into a complex interaction graph.}
\label{fig-graph}
\vspace{-4mm}
\end{figure}

\begin{definition}
\B{Complex Interaction Graph.}
Given a protein-ligand complex as shown in Figure \ref{fig-graph}(a), we define the atom node sets of protein and ligand as $\proteinV=\{a^P_1,...,a^P_m\}$ and $\ligandV=\{a^L_1,...,a^L_n\}$ with the position matrix $\proteinM \in \mathbb{R}^{m \times 3}$ and $\ligandM \in \mathbb{R}^{n \times 3}$ for 3D atomic coordinates, respectively. Then we define the complex interaction graph as a directional graph $\graph=<\mathcal{V},\mathcal{E}>$, where the vertex set $\mathcal{V} \subseteq \proteinV \cup \ligandV$ and the edge set $\mathcal{E}=f_e(\proteinV, \ligandV, \proteinM, \ligandM)$ are\hide{is} constructed based on the spatial positions of atoms\hide{complex}. The graph construction process is introduced in Appendix \ref{a-graph-constrcut}.


\end{definition}

\begin{definition}
\B{Edge-oriented Neighbors.} Given an atom node $a_i$ or a directed edge $e_{ij}$ (i.e., $a_i \rightarrow a_j$) in the complex interaction graph \graph, the edge-oriented neighbors $\mathcal{N}_{e}$ of $a_i$ or $e_{ij}$ are defined as the sets of directed edges $\{e_{ki},...,e_{li}\}$ which point to the target atom $a_i$ or the target edge $e_{ij}$.
\end{definition}
Taking Figure \ref{fig-graph}(b) as an example, the edges $e_{21}$ and $e_{41}$ are connected to the edge $e_{13}$ via the common node $a_1$, the edge-oriented neighbors of $e_{13}$ are denoted as $\mathcal{N}_{e}(e_{13})=\{e_{21},e_{41}\}$. Similarly, the edges $e_{13}$, $e_{53}$ and $e_{63}$ point to the atom node $a_3$, resulting in the neighbors set $\mathcal{N}_{e}(a_{3})=\{e_{13},e_{53},e_{63}\}$.


\begin{definition}
\B{Structure-based Protein-Ligand Binding Affinity Prediction.\hide{Structure-based Prediction of Protein-Ligand Binding Affinity.}} Given a protein-ligand complex with 3D structure, i.e., the complex interaction graph \graph \ and the 3D position matrix $M$ consisting of \proteinM and \ligandM, our goal is to learn a model $f:(\graph | M) \rightarrow y$ to precisely predict the binding affinity $y$ with preserving the spatial structure.

\end{definition}


\section{Model Framework}\label{sec-model}


\begin{figure}[t]
\centering
\subfigure[All edges]{
    \label{dist-all} 
    \includegraphics[width=0.32\columnwidth]{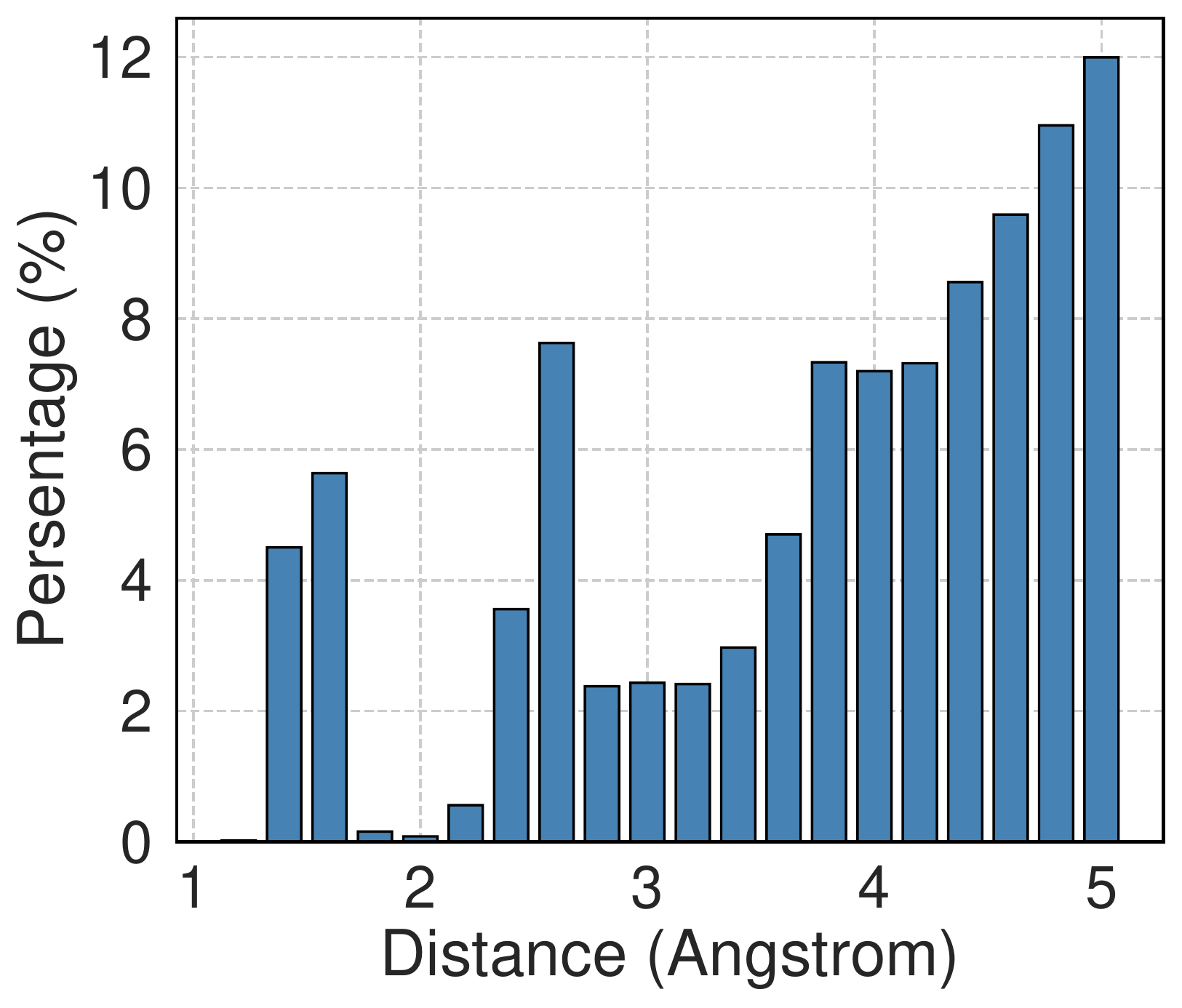}}
  \hspace{-2mm}
  \subfigure[Single bonds]{
    \label{dist-single} 
    \includegraphics[width=0.33\columnwidth]{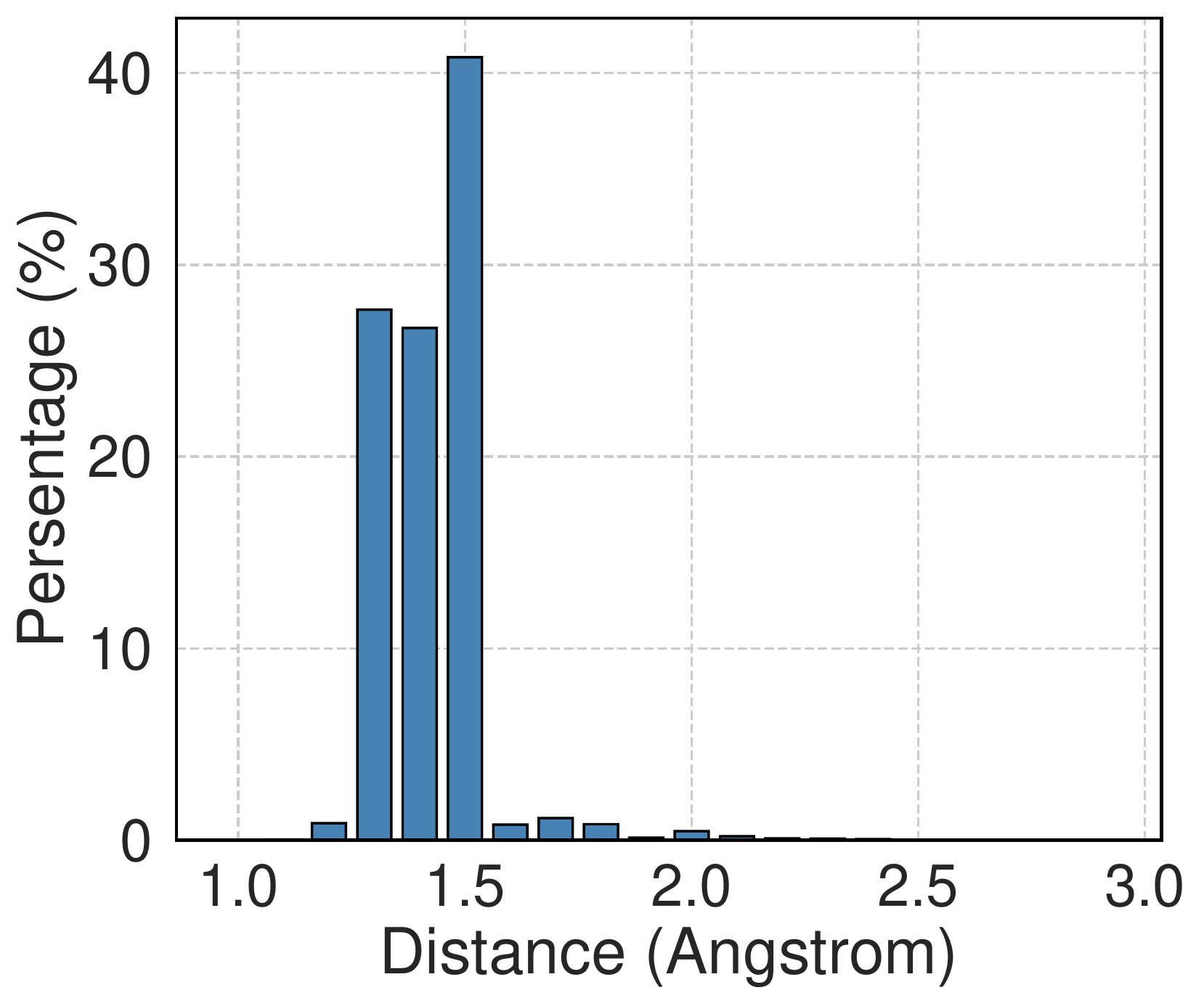}}
 \hspace{-3mm}
  \subfigure[Double bonds]{
    \label{dist-double} 
    \includegraphics[width=0.33\columnwidth]{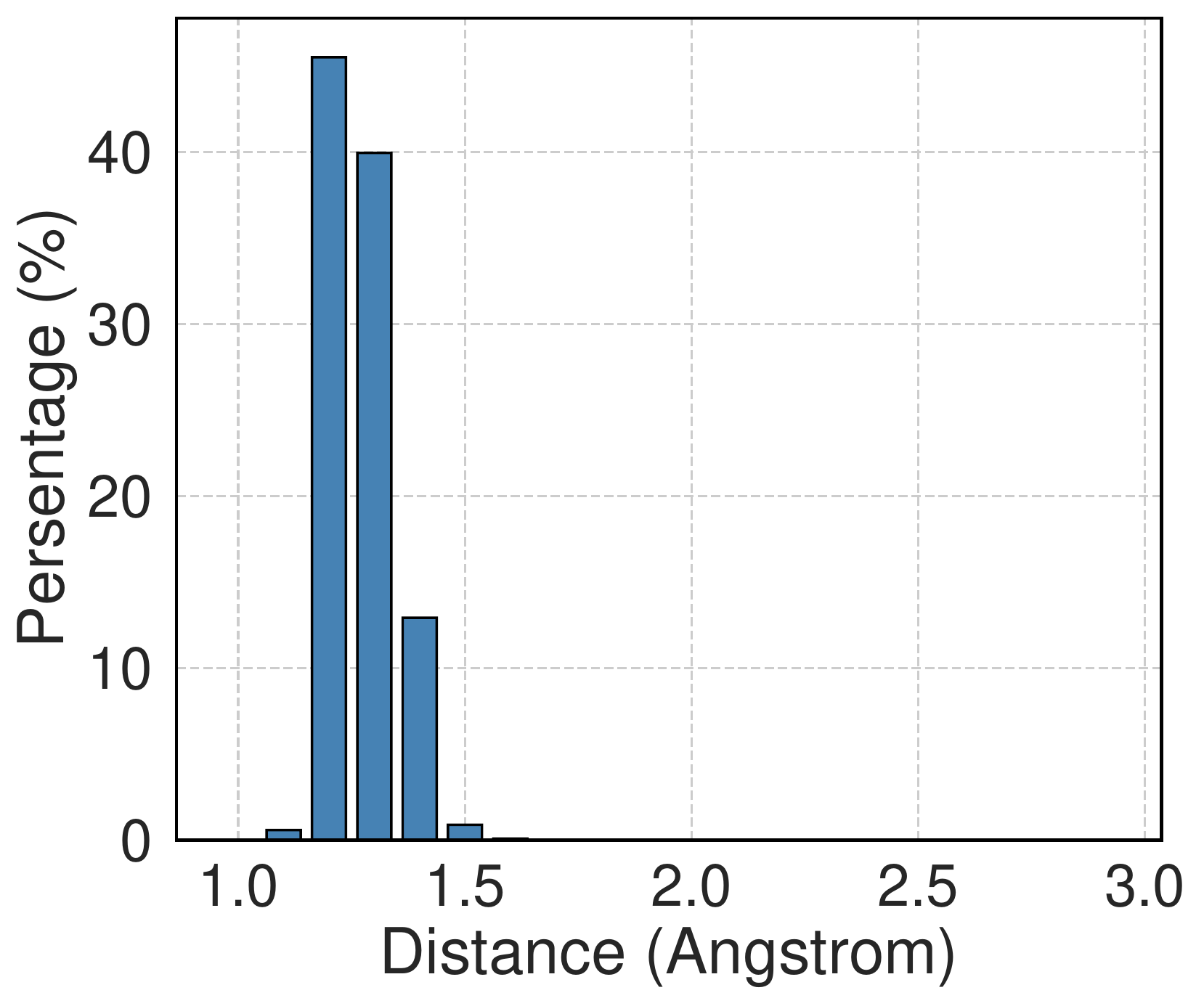}}
\vspace{-5mm}
\caption{The distribution of distance between atoms within 5 Å in the protein-ligand complex from PDBbind dataset.}
\label{fig-dist}
\vspace{-5mm}
\end{figure}

\begin{figure*}[t]
\centering
\includegraphics[width=1.\textwidth]{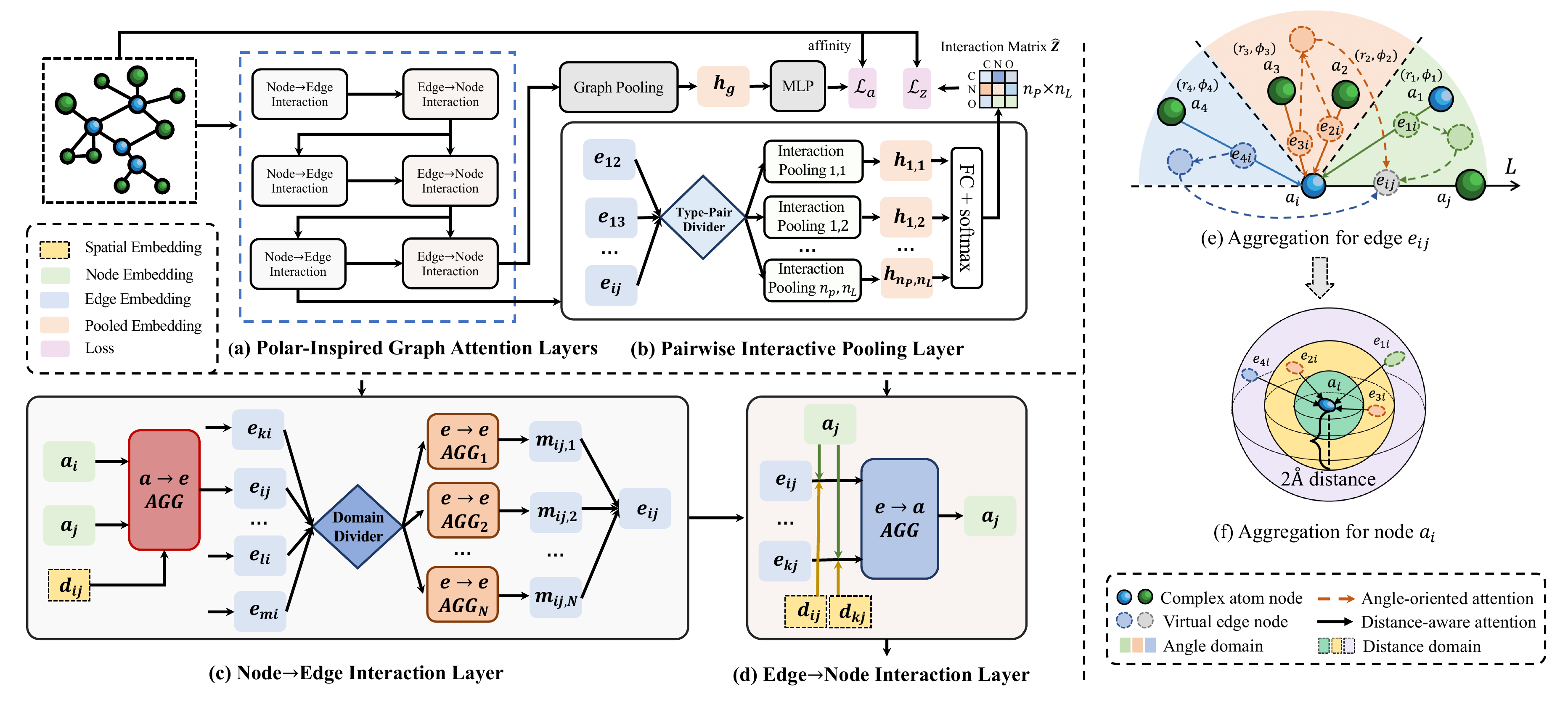}
\vspace{-7mm}
\caption{Illustration of the proposed \model framework. 
(a), (b): The two key components \gnn and \pool. (c), (d): The two inner structures of component \gnn. (e), (f): The aggregation processes in \textit{node$\rightarrow$edge} and \textit{edge$\rightarrow$node} interaction layers.
}
\label{fig-model}
\vspace{-4mm}
\end{figure*}

In this section, we present the proposed \model\hide{\footnote{Code is available at: https://github.com/agave233/SIGN}} model for protein-ligand binding affinity prediction. We first introduce the overall framework and then describe the details of each component.

\subsection{Overview}
To make accurate predictions for \dta, there are two challenges. Firstly, as shown in Figure \ref{fig-graph}, the complex graph has the unique spatial structure, which is different with general graph. Secondly, the long-range interactions between \hide{atoms of} protein and ligand are also critical to the binding affinity \cite{leckband1992long}. However, the existing GNNs are incapable of capturing such spatial information and interactions. To overcome the\hide{ above two challenges} limitations, we propose a novel \textit{\underline{S}tructure-aware \underline{I}nteractive \underline{G}raph Neural \underline{N}etwork (\model)} to model the 3D structural complex and protein-ligand spatial interactions. 

Figure \ref{fig-model} exhibits the architecture which takes the complex interaction graph \graph \ as input. We start with \hide{spatial relation embedding module for encoding the distance between atoms, then present }the polar-inspired graph attention layers (\textit{\gnn}), which are composed of \textit{node$\rightarrow$edge} and \textit{edge$\rightarrow$node} \hide{aggregation}interaction layers. \gnn can propagate the node's and edge's embeddings alternately with learning the spatial distance and angle information. The two parts of \gnn play a synergistic effect on modeling the spatial structure of the complex. After that, we apply a pairwise interactive pooling layer (\textit{\pool}) which performs on the edges' representations to obtain the atomic type-based interaction matrix of the complex. From a global view, \pool aims to approximate the overall interactions between proteins and ligands to improve the prediction performance. Finally, the model is trained through multi-task learning with augmented constraints for the interaction matrix, which serves as a self-supervised task.

\subsection{Polar Coordinate-Inspired Graph Attention}

Standard GNNs have shown great advantages in learning topological structure of the general graph, which cannot take atom's spatial position into account in the 3-dimensional space. To model the 3D structure of a complex, an intuitive method is to provide atom's 3-dimensional coordinate in the GNN architecture \cite{danel2020spatial}. However, the position information under the Cartesian coordinate system is sensitive to both translations and rotations, causing poor generalization of model when learning the complex representation. Several models, such as GNN-DTI \cite{klicpera_dimenet_2020} and MAT \cite{maziarka2020molecule}, manage to combine the distance information in the aggregation process, while only pairwise distance is not adequate. Different from DimeNet \cite{klicpera_dimenet_2020}, which specially designs Bessel functions in GNN for density functional theory (DFT) approximation with limited ability to model the larger biological complex\hide{\cite{zhang2020molecular}}, we employ iterative \textit{node$\rightarrow$edge} and \textit{edge$\rightarrow$node} interaction layers to incorporate both distance and angle information from a spatial distribution perspective.

\subsubsection{Polar-Inspired Attentive Learning Architecture.}
\label{subsec-embed}
Inspired by polar coordinate which is composed of radial distance $r$ and polar angle $\phi$, we develop an interaction-based graph attention network to leverage both the distance between nodes and the angle between edges in a collaborative framework. As illustrated in Figure \ref{fig-model}(e)\hide{\ref{fig-angle}}, \hide{when aggregating for $e_{ij}$, we treat the specific target edge $e_{ij}$ as polar axis $L$.}when aggregating for edge $e_{ij}$, we treat it as the polar axis $L$. Under such a definite polar coordinate system, the edge-oriented neighbors are distributed around $e_{ij}$ with unique identifying coordinates $(r, \phi)$. Through the method of dividing angle domains, the spatial distribution for the complex can be taken into account by means of angle-oriented attention in the first aggregation stage for edges.

Moreover, the distance factor is also helpful for structure modeling, which reveals spatial correlations. Figure \ref{fig-dist} shows the statistical distribution of distance between atoms. It can be seen that the distances of covalent bonds mainly range from 1 to 2 {\AA}, while noncovalent interactions, \hide{such as hydrogen bonds\hide{bonding}, hydrophobic interaction, and Van der Waals,}like hydrophobic and van der Waals interactions, and hydrogen bonds, are distributed over longer distances. Atomic interactions in the complex vary from different distances, which indicate different spatial relations for atom pairs. Given the radial distance $r$ between atoms $a_i$ and $a_j$, as shown in Figure \ref{fig-model}(f), we first map $r$ to a bucket (i.e., a distance domain corresponding to a type of relation) and obtain the one-hot vector $\bm{x}_{ij}$. Then we apply a dense layer transformation to get the spatial relation embedding:
\begin{equation}
\label{eq-embed}
    \bm{d}_{ij} = \bm{W}_s \bm{x}_{ij},
\end{equation}
where $\bm{W}_s \in \mathbb{R}^{d_s \times b}$ is the transformation weight matrix and $b$ is the number of buckets (i.e., spatial relations). To factor in these correlations, we design the distance-aware attention \hide{mechanism }in the second aggregation stage for nodes. As shown in Figure \ref{fig-model}(a), the overall attentive interaction process at $l$-th layer\hide{between node and edge} is defined as:
\begin{align}
    \label{eq-node-and-edge}
    \bm{e}_{ij}^{(l)} &= f^{(l)} \Big ( \big\{(\bm{a}_k^{(l-1)}, \bm{a}_i^{(l-1)}), \forall e_{ki} \in \mathcal{N}_{e}(e_{ij}) \big\} \Big ),
    \\
    \bm{a}_{j}^{(l)} &= g^{(l)} \Big ( \bm{a}_j^{(l-1)}, \big\{\bm{e}_{kj}^{(l)}, \forall e_{kj} \in \mathcal{N}_{e}(a_{j}) \big\} \Big ),
\end{align}
where $\bm{e}_{ij}^{(l)}$ is the edge embedding, $\bm{a}_{j}^{(l)}$ is the node (atom) embedding, $f(\cdot)$ and $g(\cdot)$ are interaction functions of \textit{node$\rightarrow$edge} and \textit{edge$\rightarrow$node} layers, $\mathcal{N}_{e}(e_{ij})$ and $\mathcal{N}_{e}(a_{j})$ are the edge-oriented neighbors of edge $e_{ij}$ and node $a_j$ respectively.

\subsubsection{Angle-oriented Node$\rightarrow$Edge Interaction Layer.}


Failing to distinguish neighbor nodes from different directions in the aggregation process is \hide{\li{an obvious} \zhou{a}}a weakness of the existing GNN models. To overcome this inadequacy, we adopt an angle-oriented graph attention layer to update the edge representations with integrating spatial angle information. Since the angle exists between the two edges, as shown in Figure \ref{fig-model}(c), we first get the edge embedding through aggregating the node features:
\begin{equation}
    \label{eq-a2e}
    \bm{e}_{ij}^{(l)} = \sigma (\bm{W}_{a \rightarrow e}^{(l)} \cdot [\bm{a}_i^{(l-1)} \cat \bm{a}_j^{(l-1)} \cat \bm{d}_{ij}] ),
\end{equation}
where $\bm{W}_{a \rightarrow e}^{(l)}$ is the transformation matrix for atomic combination, the operator $\cat$ represents concatenation, and $\sigma$ is the Relu \hide{activation }function.


After obtaining the representations $\{\bm{e}_{ij}^{(l)},\bm{e}_{ki}^{(l)},..,\bm{e}_{mi}^{(l)}\}$ of edge $e_{ij}$ and its neighbors, we further separate the neighboring edges in 3-dimensional space by applying an angle-domain divider \angleD, which plays an intermediate role to assign each neighbor to the specific angle domain. For example, in Figure \ref{fig-model}(e)\hide{\ref{fig-angle}}, there are four edge-oriented neighbors $e_{1i},e_{2i},e_{3i}$ and $e_{4i}$ around the central target edge $e_{ij}$. These neighboring edges are located in three different local angle domains according to the angles between edge $e_{ij}$ and its neighbors. Given the number of angle domains $N$ (e.g., $N=3$ in Figure \ref{fig-model}(e)) and the target edge $e_{ij}$ for aggregation, \angleD \  can map each neighbor $e_{ki}$ to the located angle domain index:
\begin{equation}
\label{eq-angle-index}
    Ind_{ki} = \angleD(e_{ki},e_{ij},N) = \lceil N \cdot \frac{\phi_{kij}}{180} \rceil,
\end{equation}
where $\lceil \cdot \rceil$ denotes rounding operation to get the integer index, $\phi_{kij} \in [0, 180^{\circ}]$ is the calculated angle between edges $e_{ki}$ and $e_{ij}$. Then the subset of edge-oriented neighbors which are located in $q$-th angle domain can be defined as:
\begin{equation}
\label{eq-angle-neighbor}
    \mathcal{N}^{q}_{e}(e_{ij}) = \{e_{ki} \ | \ e_{ki} \in \mathcal{N}_{e}(e_{ij}) \land Ind_{ki}=q\}.
\end{equation}
After reorganizing the neighbors of $e_{ij}$ through divider \angleD\  based on the polar coordinate system, we then feed $N$ neighbor subsets from different angle domains into $N$ independent propagation layers to capture long-range dependencies in the complex interaction graph. Firstly, we devise the domain-specific aggregation process along edges for the $q$-th angle domain:
\begin{equation}
\label{eq-e2e}
    \bm{m}^{(l)}_{ij,q} = \sum_{e_{ki} \in \mathcal{N}^{q}_{e}(e_{ij})} \alpha^{(l)}_{ki,q} \cdot \bm{e}^{(l)}_{ki}, \ \ \  1 \le q \le N,
\end{equation}
where $\bm{m}^{(l)}_{ij,q}$ is the $q$-th local aggregated edge representation at $l$-th layer, $\alpha^{(l)}_{ki,q}$ is the attention weight of the neighboring edge $e_{ki}$ across the $q$-th angle domain. Concretely, we apply the angle-oriented attention mechanism, which first uses $attn_q^l$ function to calculate the coefficient between two edges and then adopts the softmax function for normalization:
\begin{align}
\label{eq-e2e-attn}
    attn_q^l(e_{ij},e_{ki}) &= \bm{u}^T_{l,q} \cdot tanh \big(\bm{W}^{(l)}_{e, q}\cdot [\bm{e}^{(l)}_{ij} \cat \bm{e}^{(l)}_{ki}] + \bm{b}^{(l)}_{e, q} \big),
    \\
    \alpha^{(l)}_{ki,q} &= \frac{{\rm exp}\big(attn_q^l(e_{ij},e_{ki})\big)}{\sum_{e_{ti} \in \mathcal{N}^{q}_{e}(e_{ij})} {\rm exp}\big(attn_q^l(e_{ij},e_{ti})\big)},
\end{align}
where $\bm{u}_{l,q}$, $\bm{W}^{(l)}_{e, q}$ and $\bm{b}^{(l)}_{e, q}$ are the learnable attention parameters of the specific $q$-th angle domain, and we use tanh as the nonlinear activation function.

Secondly, we combine all aggregated edge embeddings obtained from Eq. (\ref{eq-e2e}). To completely preserve the spatial information in different local angle domains, we concatenate the representations as the global aggregation to update the angle-aware edge embedding:
\begin{equation}
\label{eq-e-combine}
    \bm{e}^{(l)}_{ij} = [\bm{m}^{(l)}_{ij,1} \cat \bm{m}^{(l)}_{ij,2} \cat \cdots \cat \bm{m}^{(l)}_{ij,N}].
\end{equation}

\subsubsection{Distance-aware Edge$\rightarrow$Node Interaction Layer.}
After injecting the angle information into the edge embedding $\bm{e}^{(l)}_{ij}$, we make further efforts to develop an attention-based edge$\rightarrow$node interaction layer to incorporate another spatial factor in the polar coordinate system, that is distance. As we stated in Section \ref{subsec-embed}, the distance between atoms is implicated in different meaningful correlations. Therefore, it's momentous to explore the influence of distance while learning representations for protein-ligand complexes. Specifically, since edges and nodes (atoms) have different feature spaces, \old{we first convert the edge embedding and node embedding into the hidden representation $\tilde{\bm{e}}^{(l)}_{ij}$ and $\tilde{\bm{a}}^{(l-1)}_{j}$ in the same vector space:}
\begin{align}
\label{eq-node-trans}
    \tilde{\bm{e}}^{(l)}_{ij} &= \bm{W}^{(l)}_e \cdot \bm{e}^{(l)}_{ij},
    \\
    \tilde{\bm{a}}^{(l)}_{j} &= \bm{W}^{(l)}_a \cdot \bm{a}^{(l-1)}_{j},
\end{align}
where $\bm{W}^{(l)}_e$ and $\bm{W}^{(l)}_a$ are linear transformation matrices, $\bm{a}^{(l-1)}_{j}$ is the embedding of atom $a_j$ from $(l-1)$-th layer. 

\hide{Not only because of atomic attributes, but the variant distances}As a result of the variant distances and atomic attributes, the neighboring edges have different impacts on the target \hide{atomic }node. However, the existing GNN models cannot effectively capture the influence of the distance factor. Hence, as shown in Figure \ref{fig-model}(d) and \ref{fig-model}(f), we propose to extend the original GAT \cite{velivckovic2018graph} with the distance-aware attention \old{to fuse the distance information with the capability of discriminating multiple spatial relations among atoms:}
\begin{align}
\label{eq-e2a-attn}
    w^{(l)}_{ij} &= LeakyRelu(\bm{v}^T_{l} \cdot [\tilde{\bm{e}}^{(l)}_{ij} \cat \tilde{\bm{a}}^{(l)}_{j} \cat \bm{W}^{(l)}_d \bm{d}_{ij}]),
    \\
    \beta^{(l)}_{ij} &= \frac{{\rm exp}(w^{(l)}_{ij})}{\sum_{e_{tj} \in \mathcal{N}_{e}(a_{j})} {\rm exp}(w^{(l)}_{tj})},
\end{align}
where $\bm{v}_{l}$ is the parameter of edge$\rightarrow$node attention at $l$-th layer, $\bm{W}^{(l)}_d$ is the trainable parameter matrix for distance transformation, the final calculated attention weight $\beta^{(l)}_{ij}$ reflects how important the edge $e_{ij}$ is for the node $a_j$. Then \old{we develop the distance-aware attention to multi-head attention version as GAT for better stability} and apply the aggregation process from edge to node:
\begin{equation}
\label{eq-e2a}
    \bm{a}^{(l)}_{j} = \frac{1}{C} \sum_{c=1}^{C} \sum_{e_{ij} \in \mathcal{N}_{e}(a_{j})} \beta^{(l)}_{ij,c} \cdot \tilde{\bm{e}}^{(l)}_{ij,c},
\end{equation}
where $C$ is the number of independent attention heads. Due to the angle injection for edge embedding $\tilde{\bm{e}}^{(l)}_{ij,c}$ and the distance injection for attention weight $\beta^{(l)}_{ij,c}$, our proposed model can comprehensively incorporate spatial information in the complex.

After performing $L$ polar-inspired graph attention layers, we obtain the node embedding $\bm{a}^{(L)}_{j}$ for atom $a_j$ and the edge embedding $\bm{e}^{(L)}_{ij}$ between atoms $a_i$ and $a_j$.
\hide{
\begin{figure}[t]
\centering
\includegraphics[width=0.8\columnwidth]{figure/spatial-inject-cut.pdf}
\vspace{-4mm}
\caption{Injection of angle and distance information.}
\label{fig-inject}
\vspace{-4mm}
\end{figure}
}

\subsection{Pairwise Interactive Pooling Constraint}
As introduced in Appendix \ref{a-graph-constrcut}, the constructed complex graph \graph \ only contains the partial protein structure due to the limitation of graph size and needless noise. However, the long-range intermolecular interactions between protein and ligand have\hide{a beneficial} effects on the binding affinity \cite{ballester2010machine, leckband1992long}, while \graph \ cannot provide such interactive information. To capture the \hide{distant-range}long-range interactions in the complex (e.g., the Carbon-Carbon co-occurrence interaction), we design an atomic type-aware pooling layer for edges between the protein and the ligand, which generates a proximity interaction matrix of atom type pair and enhances the representation learning process through the additional self-supervised training. 

Specifically, we first construct the pairwise interaction matrix $\bm{Z} \in \mathbb{R}^{|S_P| \times |S_L|}$ from the complete protein and its ligand, where $S_P$ and $S_L$ are atomic type sets of the protein and its ligand. Each element $T_k$ in $S_P$ or $T_l$ in $S_L$ represents the atomic number (e.g., 6) of a certain atom (e.g., carbon atom C). Following the previous work \cite{ballester2010machine}, we calculate the number of occurrences for a specific atomic type pair $(T_k,T_l)$ (e.g., (6, 7) for <C,N> pair) within a certain distance and normalize the result to get the matrix $\bm{Z}$:
\begin{align}
\label{eq-im1}
    n(T_k, T_l) &= \sum_{a_i \in \proteinV}\sum_{a_j \in \ligandV}\delta(\tau(a_i), T_k) \delta(\tau(a_j),T_l) \Theta(d_{\rho} - d_{ij}),
    \\
\label{eq-im2}
    \bm{Z}_{kl} &= \frac{n(T_k, T_l)}{\sum_{(a_i,a_j) \in \proteinV \times \ligandV}\Theta(d_{\rho} - d_{ij})},
\end{align}
where the function $\tau(a_i)$ returns the atomic type of $a_i$, $\delta(\cdot, \cdot)$ is a Kronecker delta function which outputs 1 only if the type of atom is $T_k$ (or $T_l$) and 0 otherwise, $d_{\rho}$ is referred to as the \hide{long }interaction cutoff distance and a Heaviside step function $\Theta$ is adopted to count protein–ligand atomic type pairs within the distance $d_{\rho}$. 

Secondly, we take the edge embeddings obtained from \gnn as input to the atomic type-aware pooling layer, which is shown in Figure \ref{fig-model}(b). There are $|S_P|\times|S_L|$ pooling blocks for type pairs. One block to gather edge representations belonging to atomic type pair $(T_k,T_l)$ can be formulated as:
\begin{equation}
\label{eq-pool}
   \bm{h}_{k,l} = \sum_{e_{ij} \in \mathcal{E}_I}\underbrace{ \delta(\tau(a_i), T_k) \delta(\tau(a_j),T_l)}_{{}{\rm Divider}} \bm{W}_h \bm{e}^{(L)}_{ij},
\end{equation}
where $\bm{W}_h$ is the shared parameter matrix for edge pooling, $\mathcal{E}_I \subset \mathcal{E}$ contains all the intermolecular edges in the complex $G_I$, $a_i$ and $a_j$ are atom nodes connected by $e_{ij}$, the two $\delta(\cdot,\cdot)$ functions act as a divider to pick up the corresponding edges. Then we calculate each value of the approximate interaction matrix:
\begin{equation}
\label{eq-im-normal}
   \tilde{\bm{Z}}_{kl} = \frac{{\rm exp}(\bm{q}^T \bm{h}_{k,l})}{\sum_{i,j}{\rm exp}(\bm{q}^T\bm{h}_{i,j})},
\end{equation}
where $\bm{q}$ is the trainable parameter. In the training stage, we use an additional proximity loss to draw the interaction matrix $\tilde{\bm{Z}}$ and $\bm{Z}$ closer:
\begin{equation}
    \label{eq-inter-loss}
    \bm{\mathcal{L}}_z =\sum_{\graph \in \mathcal{D}}\cat F(\tilde{\bm{Z}}) - F(\bm{Z}) \cat,
\end{equation}
where $F(\cdot)$ is the flatten operation for matrix, $\mathcal{D}$ is the training set.



\subsection{Optimization Objective}
At the last part, we add together node (atom) embeddings to get the complex representation and use MLP layers as the regressor to predict the protein-ligand binding affinity:
\begin{equation}
\label{eq-predict}
    \hat{y} = MLP\Big(\sum_{a_i \in \mathcal{V}} \bm{a}^{(L)}_i\Big).
\end{equation}

Then the absolute error between the predicted binding affinity $\hat{y}$ and the measured ground truth $y$ is used to calculate the loss. Thus, we adopt the L1 loss function to optimize the model: 
\begin{equation}
\label{eq-pred-loss}
    \bm{\mathcal{L}}_a =\sum_{\graph \in \mathcal{D}} | \hat{y} - y |,
\end{equation}
where $\mathcal{D}$ contains all the protein-ligand complexes with binding affinities. To integrate the interaction effectiveness for better complex representation learning, we further combine with the complex interaction constraint in Eq. (\ref{eq-inter-loss}) and reach the following overall objective function:
\begin{equation}
\label{eq-add-loss}
    \bm{\mathcal{L}} = \bm{\mathcal{L}}_a + \lambda\bm{\mathcal{L}}_z,
\end{equation}
where $\lambda$ is the balancing hyper-parameter to control the strength of interaction loss. The detailed process for training the proposed \model is provided in Algorithm \ref{alg-training}.

\section{Experiments} \label{sec-exp}
\begin{table*}[t]
	\caption{Performance comparison on \textit{PDBbind core set} and \textit{CSAR-HiQ set}.}
	\vspace{-3ex}
	\label{table-main-exp}
	\centering
	\scalebox{0.83}{
	\begin{tabular}{cl|cccc|cccc}
		\toprule
		\multicolumn{2}{c|}{\multirow{2}{*}{Method}} & \multicolumn{4}{c|}{PDBbind core set} & \multicolumn{4}{c}{CSAR-HiQ set} \\
		\cmidrule{3-10}
		&   & RMSE $\downarrow$  & MAE $\downarrow$  & SD $\downarrow$  & R $\uparrow$	& RMSE $\downarrow$  & MAE $\downarrow$  & SD $\downarrow$  & R $\uparrow$  \\ \midrule 
		\multirow{3}{*}{\shortstack{ML-based \\ Methods}}
		& LR &1.675 (0.000)  &1.358 (0.000)   &1.612 (0.000)  &0.671 (0.000)  &2.071 (0.000)  & 1.622 (0.000)  &1.973 (0.000)  &0.652 (0.000) \\  
		& SVR &1.555 (0.000)  &1.264 (0.000)  &1.493 (0.000)  &0.727 (0.000)  &1.995 (0.000)  &1.553 (0.000)  &1.911 (0.000)  &0.679 (0.000) \\
		& RF-Score &1.446 (0.008)  &1.161 (0.007)  &1.335 (0.010)  &0.789(0.003)  &1.947 (0.012)  &1.466 (0.009)  &1.796 (0.020)  &0.723 (0.007)
		\\ \midrule 
		
		\multirow{2}{*}{\shortstack{CNN-based \\ Methods}}
		& Pafnucy & 1.585 (0.013)  &1.284 (0.021)  &1.563 (0.022)  &0.695 (0.011)  &1.939 (0.103)  &1.562 (0.094)  &1.885 (0.071) &0.686 (0.027) \\  
		& OnionNet &1.407 (0.034)  &1.078 (0.028)  &1.391 (0.038)  &0.768 (0.014)  &1.927 (0.071)  &1.471 (0.031)  &1.877 (0.097)  &0.690 (0.040)
		\\ \midrule 
		
		\multirow{4}{*}{\shortstack{GraphDTA \\ Methods}}
		& GCN &1.735 (0.034)  &1.343 (0.037)  &1.719 (0.027)  &0.613 (0.016)  &2.324 (0.079)  &1.732 (0.065)  &2.302 (0.061)  &0.464 (0.047) \\  
		& GAT &1.765 (0.026)  &1.354 (0.033)  &1.740 (0.027)  &0.601 (0.016)  &2.213 (0.053)  &1.651 (0.061)  &2.215 (0.050)  &0.524 (0.032) \\
		& GIN &1.640 (0.044)  &1.261 (0.044)  &1.621 (0.036)  &0.667 (0.018)  &2.158 (0.074)  &1.624 (0.058)  &2.156 (0.088)  &0.558 (0.047) \\
		& GAT-GCN &1.562 (0.022)  &1.191 (0.016)  &1.558 (0.018)  &0.697 (0.008)  &1.980 (0.055)  &1.493 (0.046)  &1.969 (0.057)  &0.653 (0.026)
		\\ \midrule 
		
		\multirow{5}{*}{\shortstack{GNN-based \\ Methods}}
		& SGCN &1.583 (0.033)  &1.250 (0.036)  &1.582 (0.320)  &0.686 (0.015)  &1.902 (0.063)  &1.472 (0.067)  &1.891 (0.077)  &0.686 (0.030) \\  
		& GNN-DTI &1.492 (0.025)  &1.192 (0.032)  &1.471 (0.051)  &0.736 (0.021)  &1.972 (0.061)  &1.547 (0.058)  &1.834 (0.090)  &0.709 (0.035) \\
		& DMPNN &1.493 (0.016)  &1.188 (0.009)  &1.489 (0.014)  &0.729 (0.006)  &1.886 (0.026)  &1.488 (0.054)  &1.865 (0.035)  &0.697 (0.013) \\
		& MAT &1.457 (0.037)  &1.154 (0.037)  &1.445 (0.033)  &0.747 (0.013)  &1.879 (0.065)  &1.435 (0.058)  &1.816 (0.083)  &0.715 (0.030) \\
		& DimeNet &1.453 (0.027)  &1.138 (0.026)  &1.434 (0.023)  &0.752 (0.010)  &1.805 (0.036)  &1.338 (0.026)  &1.798 (0.027)  &0.723 (0.010) \\
		& CMPNN &1.408 (0.028)  &1.117 (0.031)  &1.399 (0.025)  &0.765 (0.009)  &1.839 (0.096)  &1.411 (0.064)  &1.767 (0.103)  &0.730 (0.052)
		\\ \midrule 
		Ours & \model &\textbf{1.316 (0.031)}  &\textbf{1.027 (0.025)}  &\textbf{1.312 (0.035)}  &\textbf{0.797 (0.012)}  &\textbf{1.735 (0.031)}  &\textbf{1.327 (0.040)}  &\textbf{1.709 (0.044)}  &\textbf{0.754 (0.014)}
		\\ \bottomrule 

	\end{tabular}}
	\vspace{-4ex}
\end{table*}
\begin{figure*}
\setlength{\abovecaptionskip}{2.mm}
\setlength{\belowcaptionskip}{-0.cm}
  \centering
  \subfigure{
    \label{general-exp-rmse} 
    \includegraphics[width=.98\columnwidth]{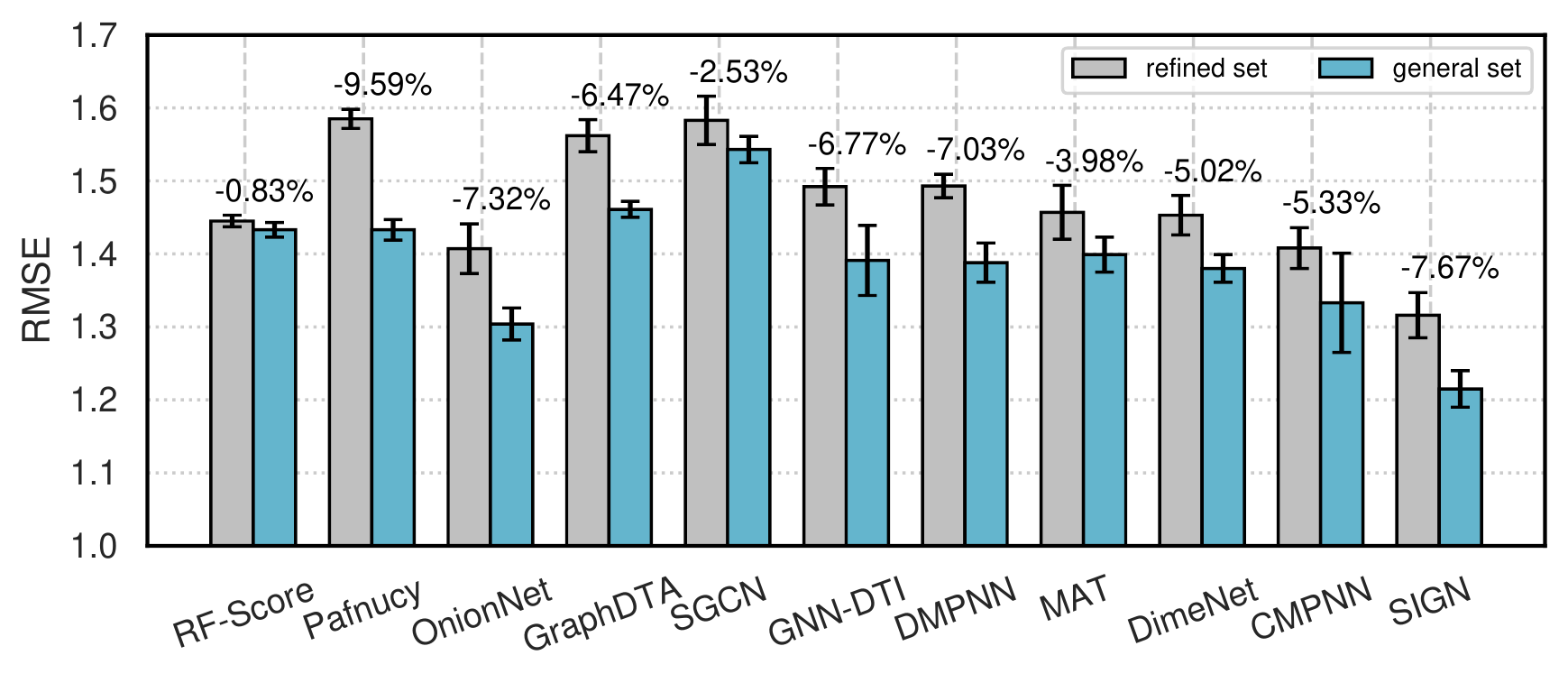}}
  \subfigure{
    \label{general-exp-mae} 
    \includegraphics[width=.98\columnwidth]{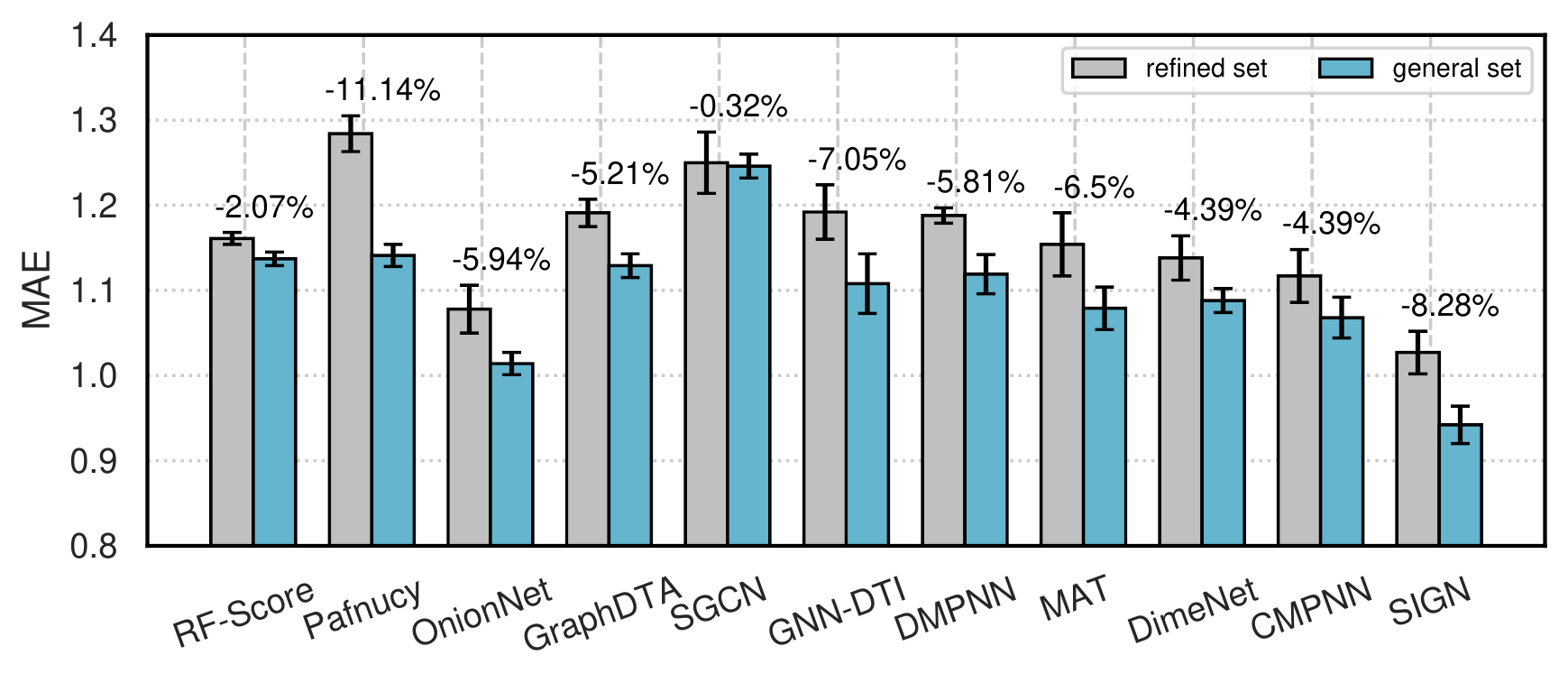}}
  \vspace{-5mm}
  \caption{Performance improvements on PDBbind benchmark when training on \textit{general set}.}
  \vspace{-5mm}
  \label{fig-general} 
\end{figure*}
In this section, we conduct experiments on two standard datasets\hide{ to evaluate our proposed model which aim} to investigate the following research questions:
\begin{itemize}[leftmargin=*,topsep=3pt]
    \item \B{RQ1.} How does the proposed \model model perform compared against the state-of-the-art methods?
    \item \B{RQ2.} How does the generalizability of \model and competitors when trained\hide{training} on the larger but lower-quality dataset? \hide{variable datasets with different scales?}
    \item \B{RQ3.} Do the spatial and interactive factors benefit the prediction?
    \item \B{RQ4.} How do the parameter settings (e.g., the cutoff distance and angle domain divisions) affect the prediction result?
\end{itemize}

\setlength{\skip\footins}{3mm}
\subsection{Experiment Settings}
\subsubsection{Datasets.}
We evaluate all models on the following \hide{publicly}public standard datasets for protein-ligand binding affinity prediction.
    

\B{PDBbind\footnote{http://www.pdbbind-cn.org}} is a well-known public dataset \cite{wang2005pdbbind} in development which provides 3D binding structures of protein-ligand complexes with experimentally determined binding affinities (refer to Appendix \ref{a-pka}). In our experiment, we mainly use the PDBbind v2016 dataset, which is most frequently used in recent works \cite{stepniewska2018development,zheng2019onionnet}. Specifically, it includes three overlapping subsets, i.e., \textit{general}, \textit{refined} and \textit{core set}. The \textit{general set} contains all 13,283 protein-ligand complexes, while the 4,057 complexes in \textit{refined set} are selected out of the \textit{general set} with better quality. Moreover, the \textit{core set} with 290 complexes serves as the highest quality\hide{CASF-2016} benchmark for testing through a careful selection process \cite{su2018comparative}. Conveniently, we call the difference between the \textit{refined} and \textit{core} subsets, that is 3,767 complexes, as \textit{refined set} of PDBbind in the following.
    
\B{CSAR-HiQ\footnote{http://www.csardock.org}} is an additional benchmark dataset \cite{dunbar2011csar}, containing two subsets with 176 and 167 protein-ligand complexes. We use this external dataset from an independent source to further evaluate the generalization ability of models.
    
\subsubsection{Setup.}
\label{exp-setup}
Following \cite{ballester2010machine}, we choose the \textit{refined set} of PDBbind as our primary training data \hide{with considering the reason that}because there is considerable overlap between the full \textit{general set} and CSAR-HiQ dataset. We\hide{first} randomly split the protein-ligand complexes in \textit{refined set} with a ratio of 9:1 for training and validation. For testing sets, we use the \textit{core set} and CSAR-HiQ \hide{data}set with removing the complexes present in \textit{refined set}.

Since the lower-quality data of \textit{general set} can still improve the performance of models \cite{li2015low}, we conduct the supplemental experiment on the full \textit{general set} which is larger but of worse quality to analyze the generalizability of our model. As stated above, we can only evaluate the performance on the \textit{core set} due to the overlapping problem of CSAR-HiQ dataset. Following \cite{stepniewska2018development,zheng2019onionnet}, we randomly select 1,000 complexes from \textit{refined set} as the validating set. The remaining 11,993 complexes in \textit{general set} are used for training\hide{as training set}.
\hide{
The statistics of datasets are summarized in Table \ref{table-dataset}.
\begin{table}[t]
	\caption{Statistical sizes of datasets.}
	\vspace{-3mm}
	\label{table-dataset}
	\centering
	\begin{tabular}{cccccc}
		\toprule
		\textbf{Dataset} & \textbf{training} & \textbf{validation} & \textbf{test (core)} & \textbf{test (csar)} \\
		\midrule
		\textit{refined set}	&	3390  &	377	&	290	&	104	\\
		\textit{general set}	&	11993	&	1000	&   290	&	/	\\
		\bottomrule
	\end{tabular}
	\vspace{-5mm}
\end{table}
}
%
\subsubsection{Evaluation Metrics.}
\old{
\hide{
To comprehensively evaluate the model performance, following \cite{stepniewska2018development, zheng2019onionnet}, we use Root Mean Square Error (RMSE) and Mean Absolute Error (MAE) to measure the prediction error. The performance of a model is also quantitatively evaluated by the classic Pearson's correlation coefficient (R) and the standard deviation (SD) in regression to measure the linear correlation between predictions and the experimental binding constants. The detail is introduced in Appendix \ref{a-implement}.
}
To comprehensively evaluate the model performance, following \cite{stepniewska2018development, zheng2019onionnet}, we use Root Mean Square Error (RMSE), Mean Absolute Error (MAE), Pearson's correlation coefficient (R) and the standard deviation (SD) in regression to measure the prediction error. The detail is introduced in Appendix \ref{a-implement}.
\hide{
As introduced in \cite{stepniewska2018development}, SD is defined as \hide{follows:
\begin{equation}
    SD = \sqrt{\frac{1}{|\mathcal{D}|-1}\sum_{i=1}^{|\mathcal{D}|} [y_i - (a+b \hat{y}_i)]^2}
\end{equation}}: $SD = \sqrt{\frac{1}{|\mathcal{D}|-1}\sum_{i=1}^{|\mathcal{D}|} [y_i - (a+b \hat{y}_i)]^2}$,
where $\hat{y}_i$ and $y_i$ respectively represent the predicted and experimental value of the $i$-th complex in dataset $\mathcal{D}$, and $a$ and $b$ are the intercept and the slope of the regression line, respectively.
}
}
\vspace{-1mm}
\subsubsection{Baselines.}
We compare our proposed model with comparative methods including machine learning-based methods (\B{LR}, \B{SVR}, and \B{RF-Score} \cite{ballester2010machine}), CNN-based methods (\B{Pafnucy} \cite{stepniewska2018development} and \B{OnionNet} \cite{zheng2019onionnet}), and GNN models \B{GraphDTA} \cite{10.1093/bioinformatics/btaa921} for protein-ligand binding affinity prediction. Moreover, various state-of-the-art GNN-based models (\B{SGCN} \cite{danel2020spatial}, \B{GNN-DTI} \cite{lim2019predicting}, \B{DMPNN} \cite{yang2019analyzing}, \B{MAT} \cite{maziarka2020molecule}, \B{DimeNet} \cite{klicpera_dimenet_2020}, and \B{CMPNN} \cite{song2020communicative}) which also consider the spatial information for molecular modeling are compared to \hide{demonstrate the effectiveness}evaluate the performance of \model. The details of \hide{implementation, }experiment settings and baseline descriptions are provided \hide{below} in Appendix \ref{a-implement} and \ref{a-baseline}.

\hide{
\subsubsection{Implementation Details.}
We implement our experiments based on Pytorch, except for Pafnucy \cite{stepniewska2018development} where we use the authors’ Tensorflow implementation. We train all models on 24 Intel CPUs and a Tesla K80 GPU with 12 GB memory. Note that DimeNet \cite{klicpera_dimenet_2020} is not compared in our experiment. \zhoucom{change lager} \li{ because it can not model the larger molecules and will raise the out-of-memory error when training protein-ligand complexes.} 
The model input and parameter settings for baselines and our method are introduced in Appendix \ref{a-implement}.
}
\vspace{-1mm}

\subsection{Performance Evaluation}
\subsubsection{Overall Comparison (RQ1).}
We first compare our proposed \model with baseline approaches on two benchmark datasets. As shown in Table \ref{table-main-exp}, the average and the standard deviation of four indicators for testing performance are reported across five random runs. In general, we can observe that \model achieves the best performance on two datasets, with 6.5\% and 3.9\% improvement of RMSE over the best baseline models on PDBbind and CSAR-HiQ datasets, respectively. We further have the following observations.

Among all baselines, GraphDTA methods show relatively poor performance due to the failure of considering the spatial structure and interactions between proteins and ligands. It indicates that simply modeling the molecular graph with protein sequence information is not capable of predicting structure-based protein-ligand binding affinity. By contrast, from the perspective of interaction modeling, the machine learning-based methods and OnionNet model take advantage of long-range interaction features and achieve better results. However, these data-driven approaches relying on feature engineering ignore the informative spatial structures of complexes and have limited generalization capability on the additional CSAR-HiQ dataset.
From the perspective of spatial structural modeling, we find that SGCN and GNN-DTI which incorporate position and distance information exhibit considerable improvement over the vanilla GCN and GAT. Since SGCN takes atomic position coordinates as input directly, it will be easily affected by the rotation and translation of atoms, and the 3D CNN model Pafnucy suffers from a similar issue. \hide{As a result}Thus, the prediction results \hide{of them }are not ideal. Despite leveraging a transformer-like attention mechanism to handle the spatial structure, MAT is not better than RF-Score and OnionNet, suggesting the importance of combining spatial and interactive information. The edge-oriented model CMPNN outperforms \hide{other GNN-based}the above methods because it enhances DMPNN with communication while propagating the distance information, which shows the significance of node-edge message passing process. Although DimeNet can learn from angle information and perform slightly better\hide{better slightly}, the performance is still not ideal due to its limited ability of modeling larger biomolecules. \hide{its limited ability of modeling larger biomolecules causes unsatisfactory performance on the binding affinity prediction.}Our proposed \model can not only capture more comprehensive angle-enhanced structural information instead of \hide{only}just distance, but also handle interactions in the complex through multi-task learning framework. Therefore, \model is much effective for modeling the protein-ligand complex and can accurately predict the binding affinity.


\subsubsection{Generalizability Comparison (RQ2).}


There is increasing 3D structure-based protein-ligand data with binding affinity, whereas the amount of high-quality data in \textit{refined set} is relatively small. Thus, the ability of utilizing \hide{the }more lower-quality data to improve performance\hide{for performance improvement} shows the generalizability of model, which is another necessary measurement of performance evaluation. As introduced in Section \ref{exp-setup}, we conduct the extra experiment of generalizability on the \textit{general set} of PDBbind dataset.
As illustrated in Figure \ref{fig-general}, we compare the proposed \model with major 
competitive baselines on two training sets. The results show that \model gets the lowest prediction error remarkably under both training settings\hide{on both \textit{refined set} and \textit{general set}}. More importantly, our model improves the performance by around 8\% when \hide{training}trained on the \textit{general set} and it further expands the prediction advantage compared to baselines. Therefore, \model is proved to be more generalizable to more data in large quantity but poor quality. \hide{it demonstrates \model is more generalizable when we have access to more data with large quantity but poor quality.}

\subsubsection{Impact of Spatial and Interactive Factors (RQ3).}

\begin{figure}
\setlength{\abovecaptionskip}{2.mm}
\setlength{\belowcaptionskip}{-0.cm}
  \centering
  \subfigure{
    \label{component-exp-rmse} 
    \includegraphics[width=0.47\columnwidth]{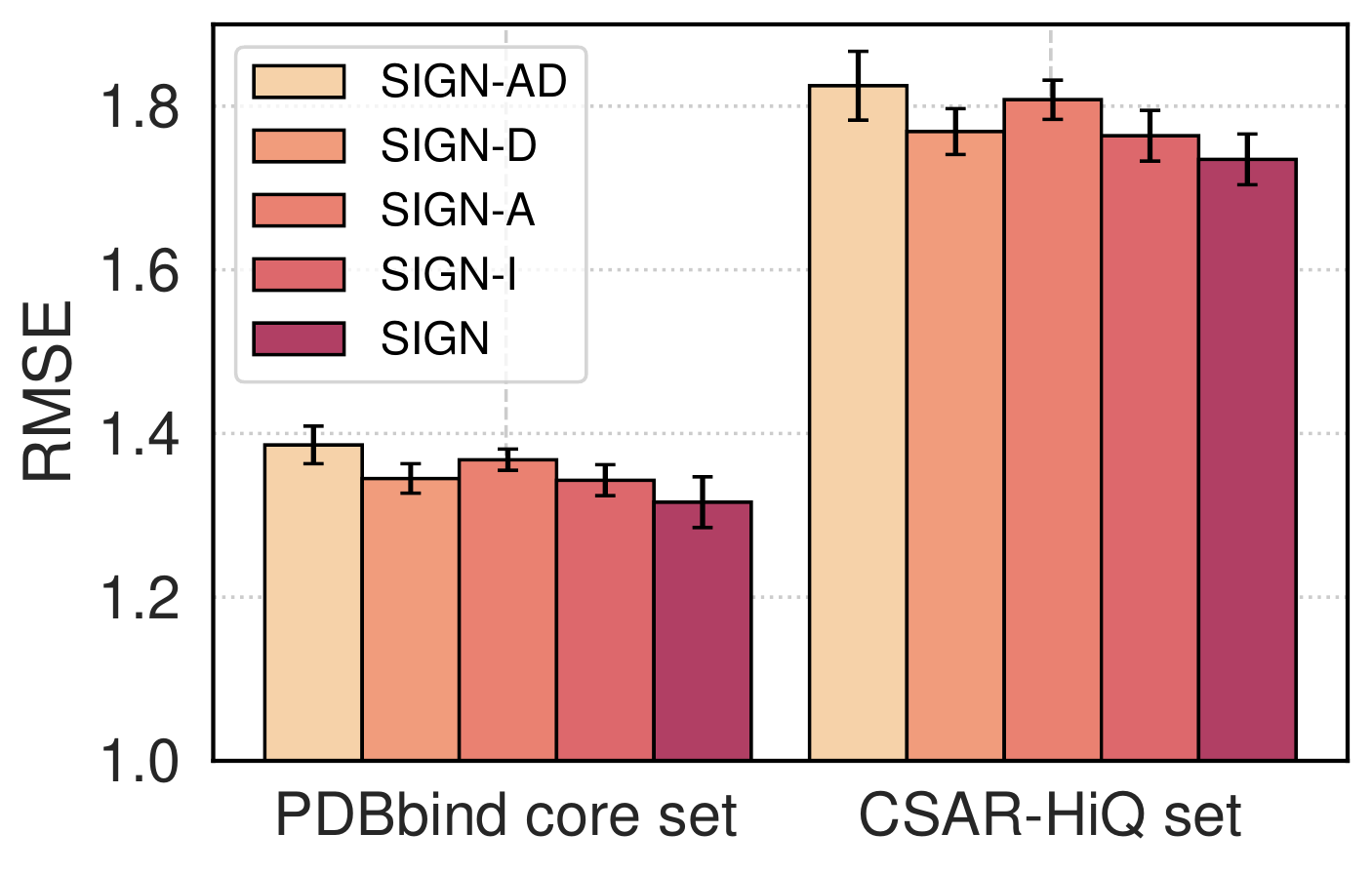}}
      \subfigure{
    \label{component-exp-mae} 
    \includegraphics[width=0.47\columnwidth]{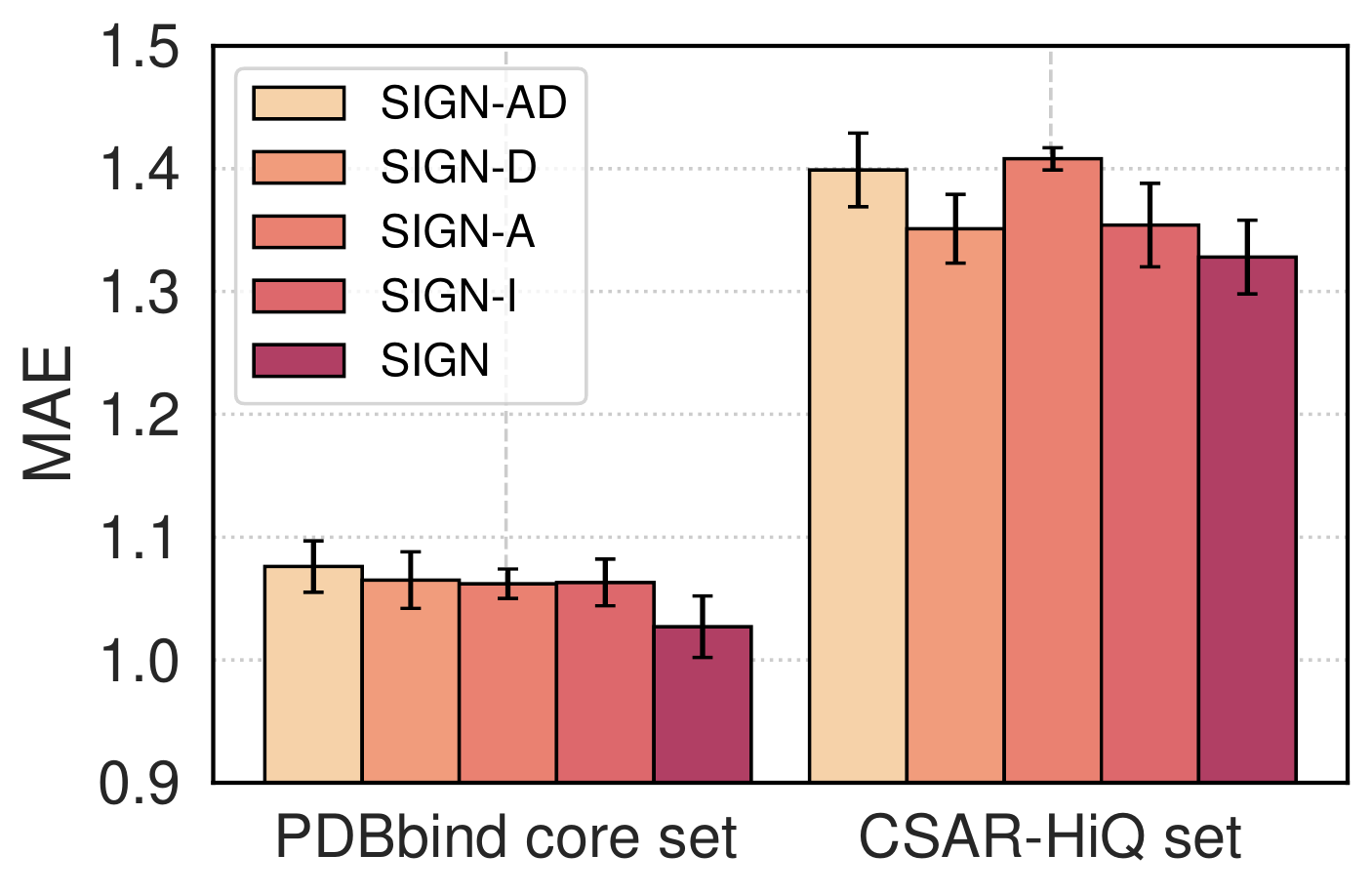}}
\\[-3ex]
  \subfigure{
    \label{component-exp-sd} 
    \includegraphics[width=0.47\columnwidth]{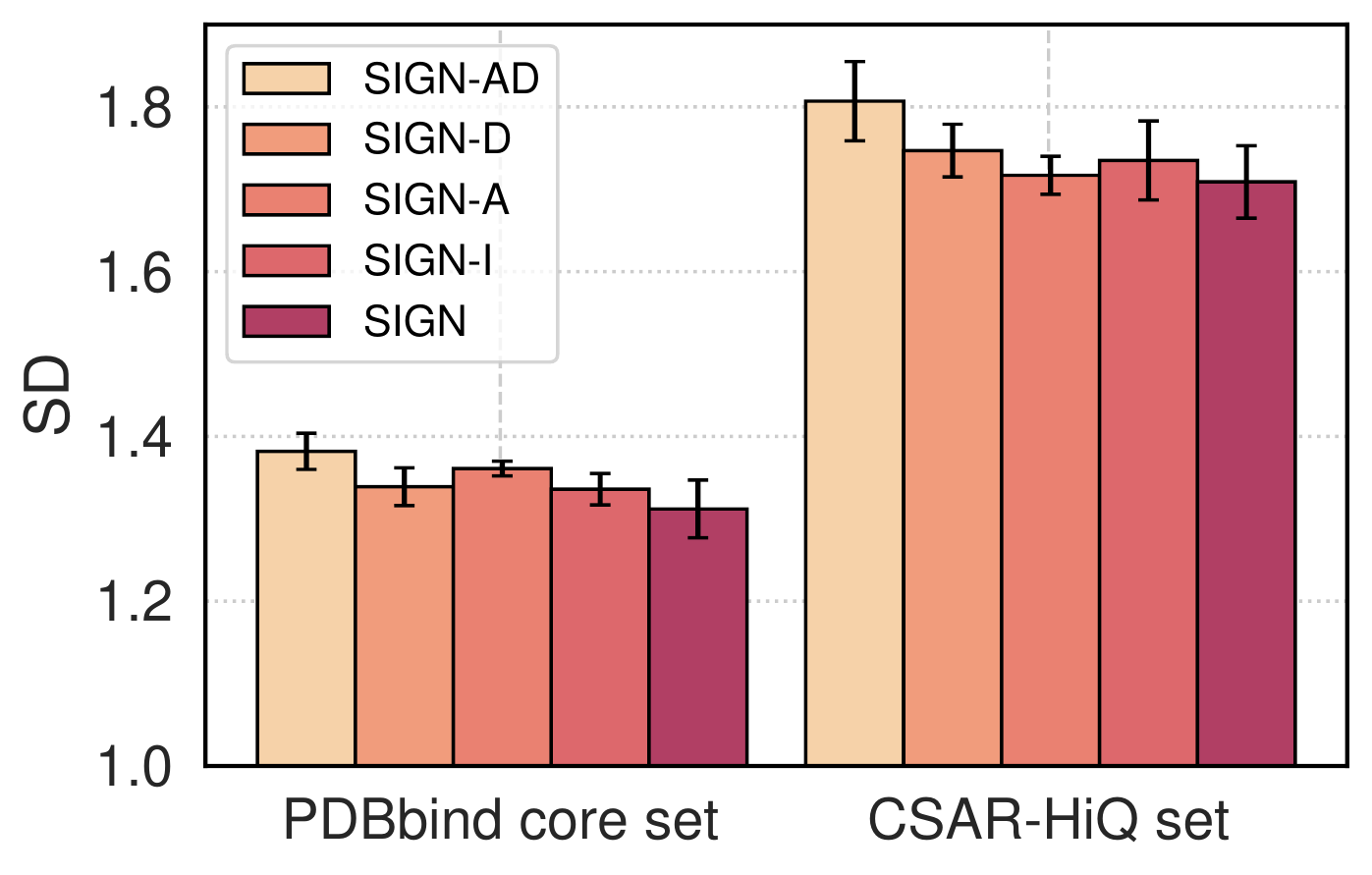}}
      \subfigure{
    \label{component-exp-r} 
    \includegraphics[width=0.47\columnwidth]{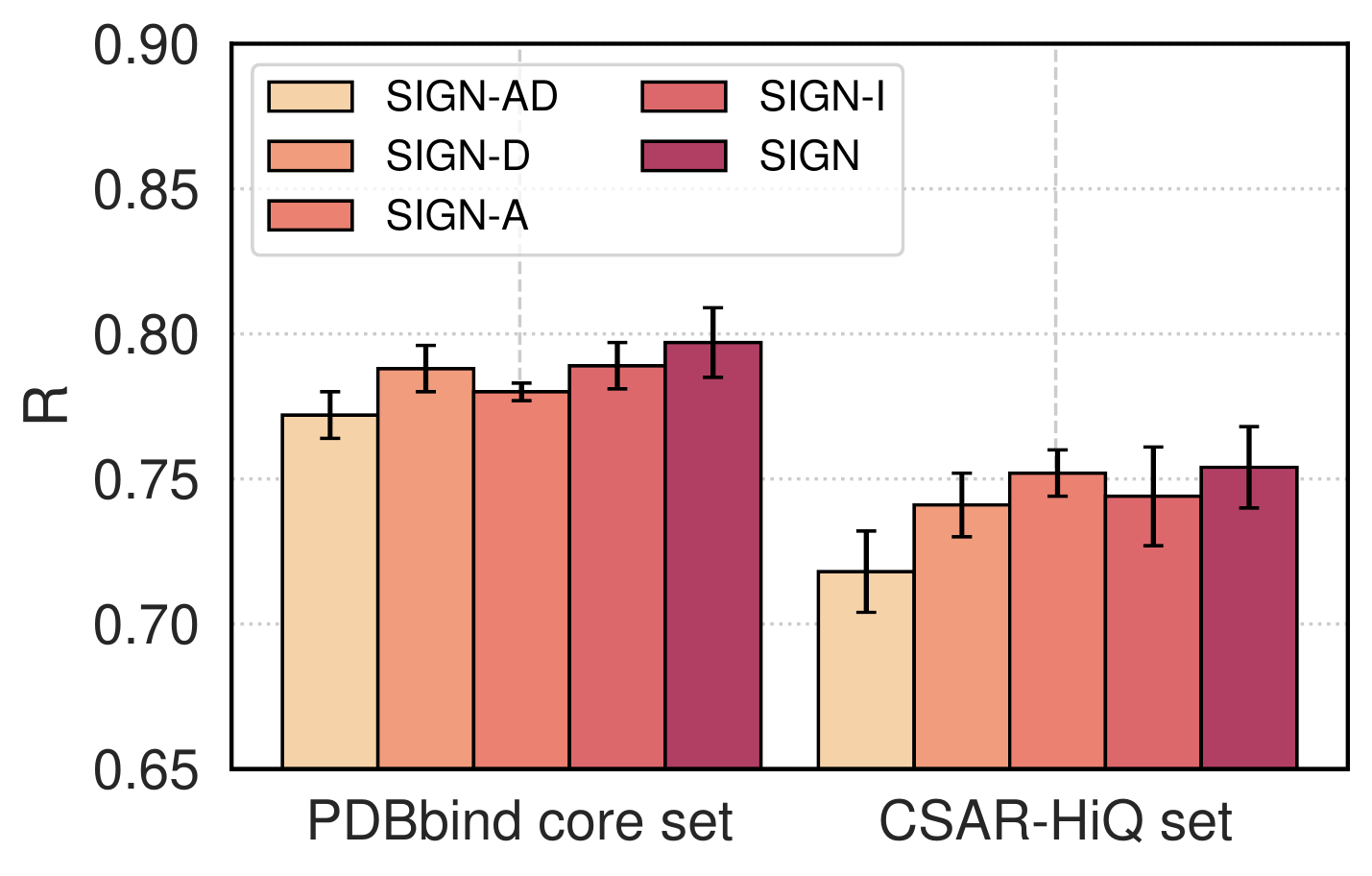}}
  \vspace{-3mm}
  \caption{Contribution of spatial and interactive factors.\hide{ for \model with its variants on two datasets.}}
  \vspace{-6mm}
  \label{fig-ablation} 
\end{figure}

To verify the effectiveness of factors that influence the final performance, we compare \model with its variants on the two benchmarks.
\begin{itemize}[leftmargin=*,topsep=1pt]
    \item \B{\textsf{\model-AD}} uses the vanilla GAT layers for node-edge interaction without either angle or distance information.
    \item \B{\textsf{\model-D}} uses the vanilla GAT layer without distance information.
    \item \B{\textsf{\model-A}} uses the vanilla GAT layer without angle information.
    \item \B{\textsf{\model-I}} removes the interaction loss $\mathcal{L}_z$.
\end{itemize}
Figure \ref{fig-ablation} presents the comparison results on all metrics. As we can see, the proposed \model \hide{can }outperforms other variants, proving the necessity of handling the spatial and interactive information synergistically which is essential for protein-ligand binding affinity prediction. Specifically, \textsf{\model-D} and \textsf{\model-A} perform worse than \model since they can only capture the one-sided spatial structural information, i.e., distance or angle information in the complex. Furthermore, the prediction error of \textsf{\model-AD} is especially high among all variants. It indicates that modeling the complete spatial structure has a significant impact on performance improvement. The lack of long-range interactions in \textsf{\model-I} also leads to performance reduction, which confirms that only utilizing the spatial factors is insufficient and will lose the important interactive information.

\subsubsection{Parameters Analysis (RQ4).}
\label{sec-para-analy}

\begin{figure}
\setlength{\abovecaptionskip}{2.mm}
\setlength{\belowcaptionskip}{-0.cm}
  \centering
  \hide{
  \subfigure{
    \label{parameter-core-layer} 
    \includegraphics[width=0.47\columnwidth]{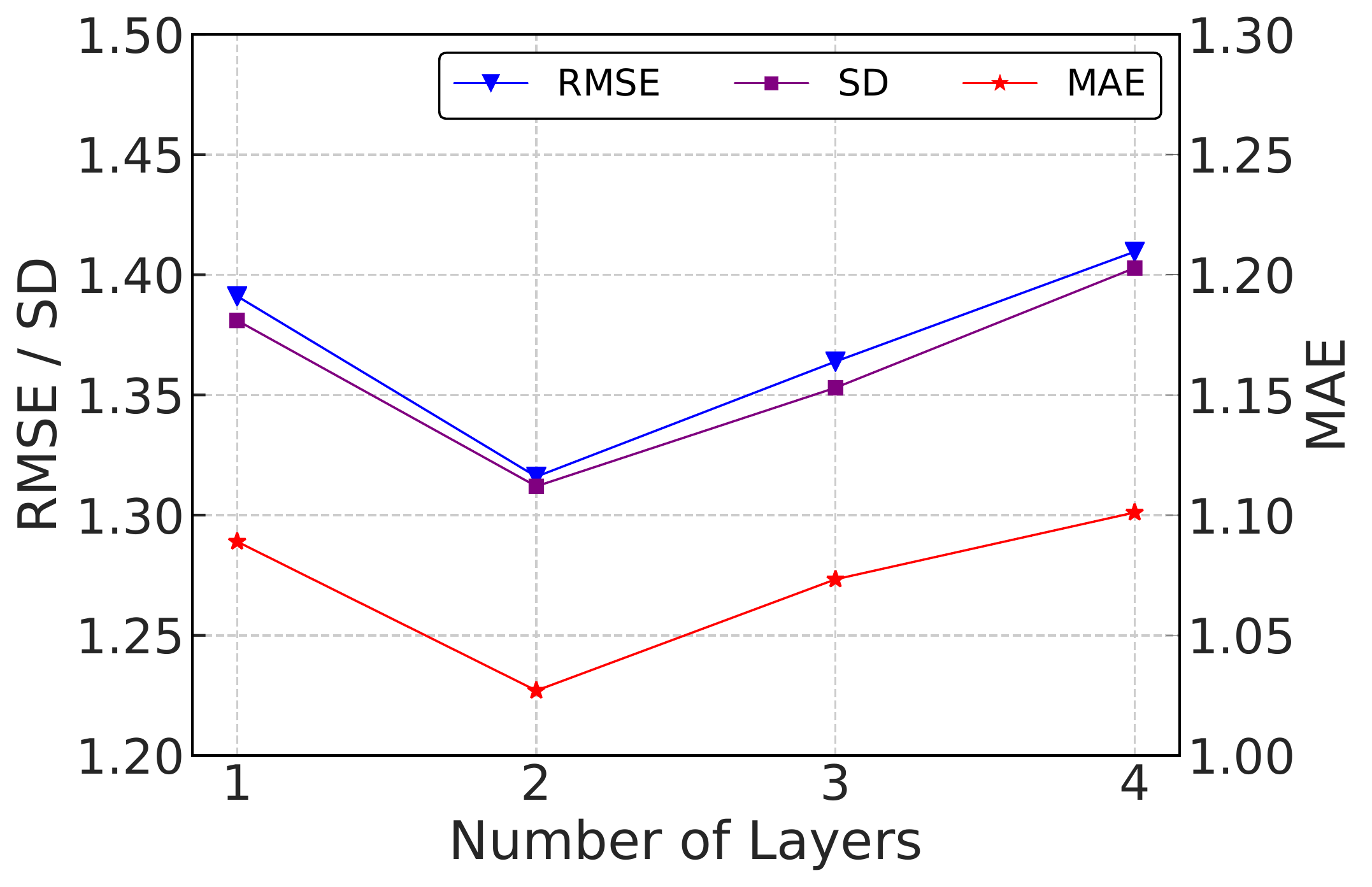}}}
      \subfigure{
    \label{parameter-core-cutoff} 
    \includegraphics[width=0.47\columnwidth]{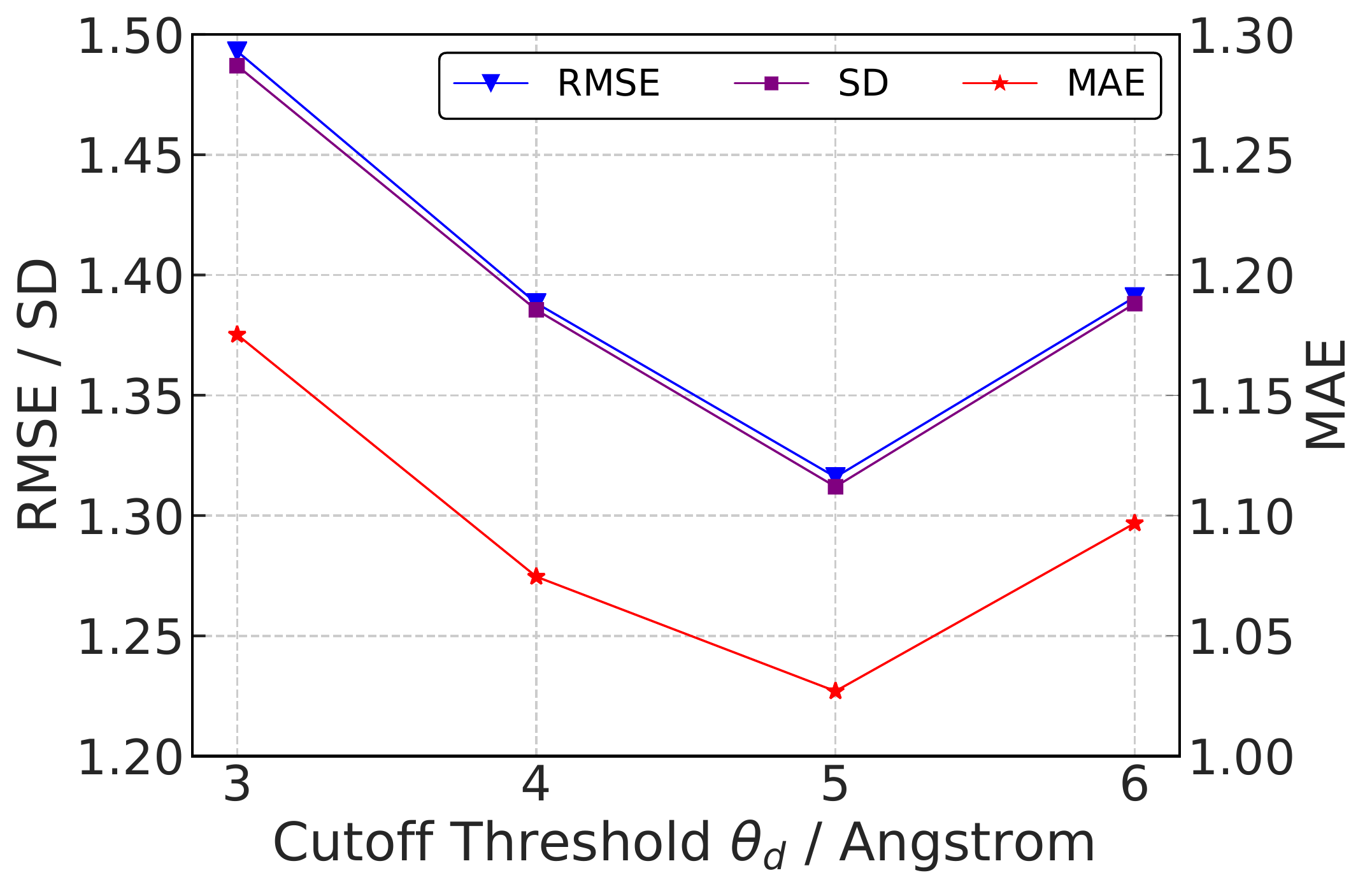}}
  \subfigure{
    \label{parameter-core-angle} 
    \includegraphics[width=0.47\columnwidth]{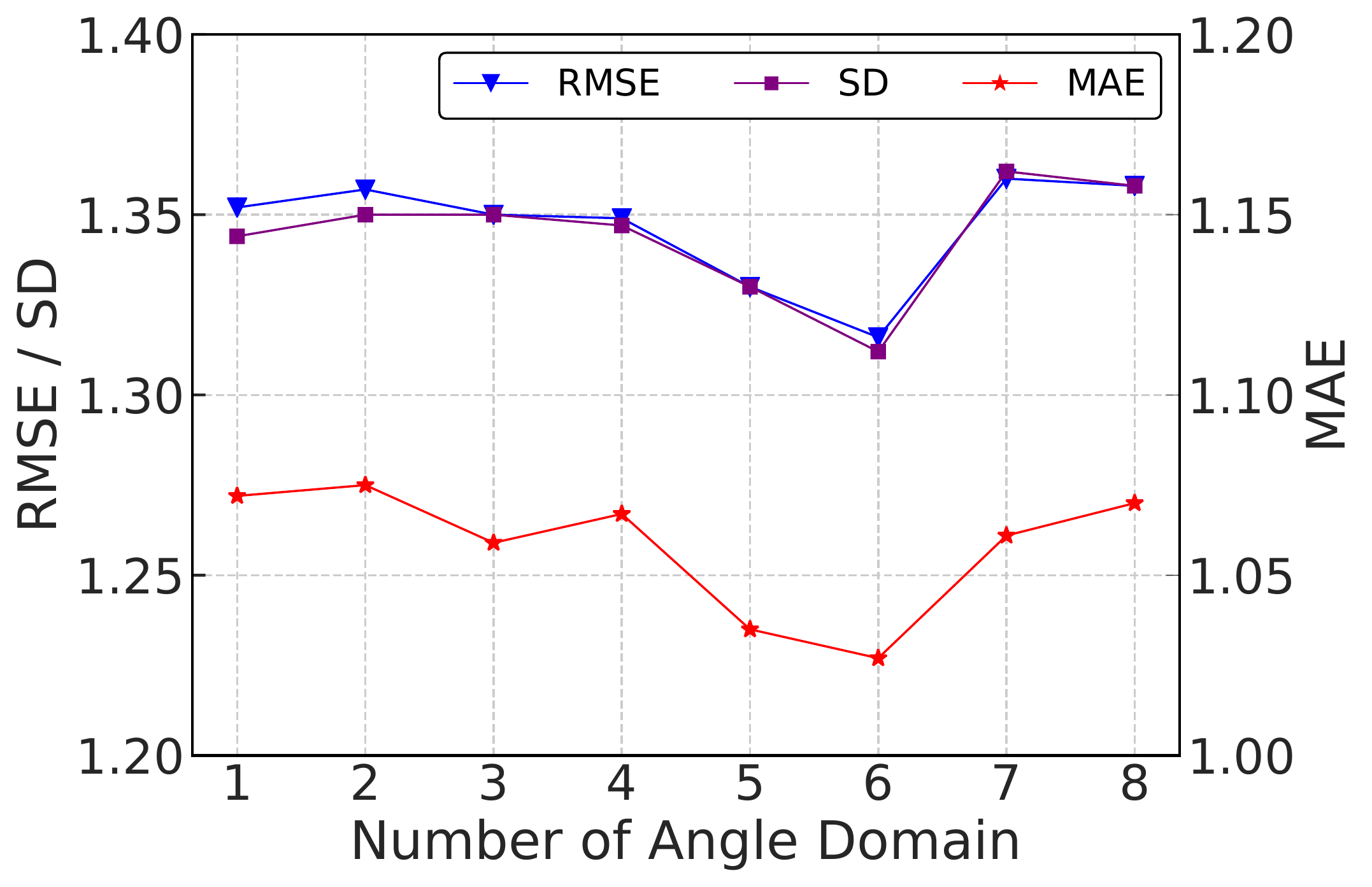}}\hide{
      \subfigure{
    \label{parameter-core-lambda} 
    \includegraphics[width=0.47\columnwidth]{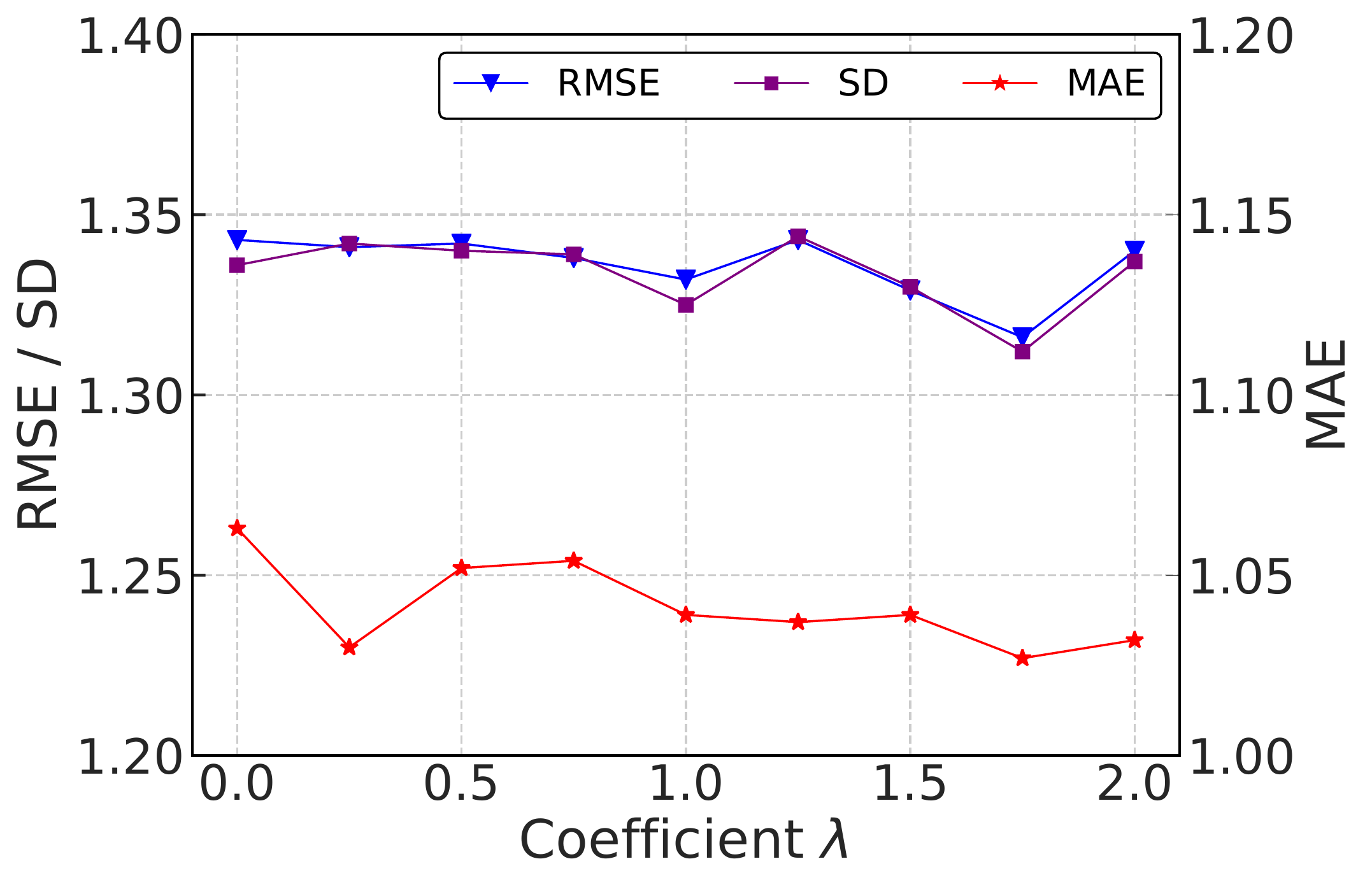}}}
  \vspace{-2mm}
  \caption{Parameter analysis on PDBbind \textit{core set}.}
  \vspace{-6mm}
  \label{fig-parameter} 
\end{figure}


As depicted in Figure \ref{fig-parameter}, we investigate the performance variation for \model w.r.t several necessary hyper-parameters by varying each parameter \hide{in a specified range} while keeping others fixed as default settings. More results are in Appendix \ref{a-para}.

\hide{
\B{Number of \gnn layers $L$}. We first study the influence of multi-hop propagation with stacking node-edge interaction layers from 1 to 4. We observe that increasing the number of layers would not always give rise to a better result. The model with one \gnn layer has limited ability to model high-order information in the complex. As a result of over-fitting, the performance of the model using more than 3 layers starts to degenerate gradually. Therefore, applying two interaction layers in \model is enough to capture sufficient spatial information.}

\B{Cutoff distance $\theta_d$}. We\hide{ then} analyze the effect of cutoff distance for complex graph construction when varying $\theta_d$ from 3 to 6. With the increase of \hide{cutoff distance}$\theta_d$, more spatial information in the complex is available to our model and beneficial for learning complex representation better\hide{better complex representation}, which leads to dramatic performance improvements when $\theta_d \le 5$ Å. After that, too long\hide{large} cutoff distance will introduce additional redundancies and degrade the performance.


\B{Angle domain divisions $N$}. To look deeper into the impact of angle information in our model, we divide the angle domains varying from 1 to 8. We can see that the model performs best when the number of angle domains is 5 or 6. Too fine-grained or coarse-grained divisions will result in performance degradation. One possible explanation is that too fine-grained divisions cannot provide distinguishable information in space while the angle domain at a too big granularity contains quite sparse atomic neighbors, both of which\hide{which both} have an adverse effect on learning spatial information.

\hide{
\B{Balancing coefficient $\lambda$}. Moreover, we change the coefficient $\lambda$ to control the trade-off between the prediction loss and interaction loss. From the results, we observe that the performance first tends to get better with incorporating more interactive information for long-range dependencies, and then begins to drop off slightly. In general, our model is stable with varying coefficients and always achieves better performance than all baseline methods.}

\vspace{-3mm}


\section{Conclusion}
In this paper, we investigated how to improve the prediction of binding affinity between proteins and ligands. Specifically, we proposed a  GNN-based model, \model, to learn the representations of protein-ligand complexes for better binding affinity prediction by leveraging the fine-grained structure and interaction information among atoms. Along this line, we designed the polar-inspired graph attention layers (\gnn) to integrate both distance and angle information for 3D spatial structure modeling. Also, to further improve the prediction performance, we introduced a well-designed pooling process along with a reconstruction learning task for interaction matrix. Finally, the experimental results on two benchmarks showed the effectiveness and the generalizability of the proposed model.
\vspace{-3mm}



\normalem
\bibliographystyle{ACM-Reference-Format}
\bibliography{ref}


\begin{thebibliography}{48}


\ifx \showCODEN    \undefined \def \showCODEN     #1{\unskip}     \fi
\ifx \showDOI      \undefined \def \showDOI       #1{#1}\fi
\ifx \showISBNx    \undefined \def \showISBNx     #1{\unskip}     \fi
\ifx \showISBNxiii \undefined \def \showISBNxiii  #1{\unskip}     \fi
\ifx \showISSN     \undefined \def \showISSN      #1{\unskip}     \fi
\ifx \showLCCN     \undefined \def \showLCCN      #1{\unskip}     \fi
\ifx \shownote     \undefined \def \shownote      #1{#1}          \fi
\ifx \showarticletitle \undefined \def \showarticletitle #1{#1}   \fi
\ifx \showURL      \undefined \def \showURL       {\relax}        \fi
\providecommand\bibfield[2]{#2}
\providecommand\bibinfo[2]{#2}
\providecommand\natexlab[1]{#1}
\providecommand\showeprint[2][]{arXiv:#2}

\bibitem[\protect\citeauthoryear{Allen, Balius, Mukherjee, Brozell, Moustakas,
  Lang, Case, Kuntz, and Rizzo}{Allen et~al\mbox{.}}{2015}]%
        {allen2015dock}
\bibfield{author}{\bibinfo{person}{William~J Allen}, \bibinfo{person}{Trent~E
  Balius}, \bibinfo{person}{Sudipto Mukherjee}, \bibinfo{person}{Scott~R
  Brozell}, \bibinfo{person}{Demetri~T Moustakas}, \bibinfo{person}{P~Therese
  Lang}, \bibinfo{person}{David~A Case}, \bibinfo{person}{Irwin~D Kuntz}, {and}
  \bibinfo{person}{Robert~C Rizzo}.} \bibinfo{year}{2015}\natexlab{}.
\newblock \showarticletitle{DOCK 6: Impact of new features and current docking
  performance}.
\newblock \bibinfo{journal}{\emph{Journal of computational chemistry}}
  \bibinfo{volume}{36}, \bibinfo{number}{15} (\bibinfo{year}{2015}),
  \bibinfo{pages}{1132--1156}.
\newblock


\bibitem[\protect\citeauthoryear{Ballester and Mitchell}{Ballester and
  Mitchell}{2010}]%
        {ballester2010machine}
\bibfield{author}{\bibinfo{person}{Pedro~J Ballester} {and}
  \bibinfo{person}{John~BO Mitchell}.} \bibinfo{year}{2010}\natexlab{}.
\newblock \showarticletitle{A machine learning approach to predicting
  protein--ligand binding affinity with applications to molecular docking}.
\newblock \bibinfo{journal}{\emph{Bioinformatics}} \bibinfo{volume}{26},
  \bibinfo{number}{9} (\bibinfo{year}{2010}), \bibinfo{pages}{1169--1175}.
\newblock


\bibitem[\protect\citeauthoryear{Batool, Ahmad, and Choi}{Batool
  et~al\mbox{.}}{2019}]%
        {batool2019structure}
\bibfield{author}{\bibinfo{person}{Maria Batool}, \bibinfo{person}{Bilal
  Ahmad}, {and} \bibinfo{person}{Sangdun Choi}.}
  \bibinfo{year}{2019}\natexlab{}.
\newblock \showarticletitle{A structure-based drug discovery paradigm}.
\newblock \bibinfo{journal}{\emph{International journal of molecular sciences}}
  \bibinfo{volume}{20}, \bibinfo{number}{11} (\bibinfo{year}{2019}),
  \bibinfo{pages}{2783}.
\newblock


\bibitem[\protect\citeauthoryear{Callaway}{Callaway}{2020}]%
        {callaway2020will}
\bibfield{author}{\bibinfo{person}{Ewen Callaway}.}
  \bibinfo{year}{2020}\natexlab{}.
\newblock \showarticletitle{'It will change everything': DeepMind's AI makes
  gigantic leap in solving protein structures.}
\newblock \bibinfo{journal}{\emph{Nature}} (\bibinfo{year}{2020}).
\newblock


\bibitem[\protect\citeauthoryear{Colwell}{Colwell}{2018}]%
        {colwell2018statistical}
\bibfield{author}{\bibinfo{person}{Lucy~J Colwell}.}
  \bibinfo{year}{2018}\natexlab{}.
\newblock \showarticletitle{Statistical and machine learning approaches to
  predicting protein--ligand interactions}.
\newblock \bibinfo{journal}{\emph{Current opinion in structural biology}}
  \bibinfo{volume}{49} (\bibinfo{year}{2018}), \bibinfo{pages}{123--128}.
\newblock


\bibitem[\protect\citeauthoryear{Danel, Spurek, Tabor, {\'S}mieja, Struski,
  S{\l}owik, and Maziarka}{Danel et~al\mbox{.}}{2020}]%
        {danel2020spatial}
\bibfield{author}{\bibinfo{person}{Tomasz Danel},
  \bibinfo{person}{Przemys{\l}aw Spurek}, \bibinfo{person}{Jacek Tabor},
  \bibinfo{person}{Marek {\'S}mieja}, \bibinfo{person}{{\L}ukasz Struski},
  \bibinfo{person}{Agnieszka S{\l}owik}, {and} \bibinfo{person}{{\L}ukasz
  Maziarka}.} \bibinfo{year}{2020}\natexlab{}.
\newblock \showarticletitle{Spatial graph convolutional networks}. In
  \bibinfo{booktitle}{\emph{International Conference on Neural Information
  Processing\hide{ICONIP}}}. Springer, \bibinfo{pages}{668--675}.
\newblock


\bibitem[\protect\citeauthoryear{Do, Tran, and Venkatesh}{Do
  et~al\mbox{.}}{2019}]%
        {do2019graph}
\bibfield{author}{\bibinfo{person}{Kien Do}, \bibinfo{person}{Truyen Tran},
  {and} \bibinfo{person}{Svetha Venkatesh}.} \bibinfo{year}{2019}\natexlab{}.
\newblock \showarticletitle{Graph transformation policy network for chemical
  reaction prediction}. In \bibinfo{booktitle}{\emph{\hide{Proceedings of the
  25th ACM SIGKDD International Conference on Knowledge Discovery \& Data
  Mining}SIGKDD}}. \bibinfo{pages}{750--760}.
\newblock


\bibitem[\protect\citeauthoryear{Dunbar~Jr, Smith, Yang, Ung, Lexa, Khazanov,
  Stuckey, Wang, and Carlson}{Dunbar~Jr et~al\mbox{.}}{2011}]%
        {dunbar2011csar}
\bibfield{author}{\bibinfo{person}{James~B Dunbar~Jr},
  \bibinfo{person}{Richard~D Smith}, \bibinfo{person}{Chao-Yie Yang},
  \bibinfo{person}{Peter Man-Un Ung}, \bibinfo{person}{Katrina~W Lexa},
  \bibinfo{person}{Nickolay~A Khazanov}, \bibinfo{person}{Jeanne~A Stuckey},
  \bibinfo{person}{Shaomeng Wang}, {and} \bibinfo{person}{Heather~A Carlson}.}
  \bibinfo{year}{2011}\natexlab{}.
\newblock \showarticletitle{CSAR benchmark exercise of 2010: selection of the
  protein--ligand complexes}.
\newblock \bibinfo{journal}{\emph{Journal of chemical information and
  modeling}} \bibinfo{volume}{51}, \bibinfo{number}{9} (\bibinfo{year}{2011}),
  \bibinfo{pages}{2036--2046}.
\newblock


\bibitem[\protect\citeauthoryear{Gohlke, Hendlich, and Klebe}{Gohlke
  et~al\mbox{.}}{2000}]%
        {gohlke2000knowledge}
\bibfield{author}{\bibinfo{person}{Holger Gohlke}, \bibinfo{person}{Manfred
  Hendlich}, {and} \bibinfo{person}{Gerhard Klebe}.}
  \bibinfo{year}{2000}\natexlab{}.
\newblock \showarticletitle{Knowledge-based scoring function to predict
  protein-ligand interactions}.
\newblock \bibinfo{journal}{\emph{Journal of molecular biology}}
  \bibinfo{volume}{295}, \bibinfo{number}{2} (\bibinfo{year}{2000}),
  \bibinfo{pages}{337--356}.
\newblock


\bibitem[\protect\citeauthoryear{Hao, Lu, Huang, Wang, Hu, Liu, Chen, and
  Lee}{Hao et~al\mbox{.}}{2020}]%
        {hao2020asgn}
\bibfield{author}{\bibinfo{person}{Zhongkai Hao}, \bibinfo{person}{Chengqiang
  Lu}, \bibinfo{person}{Zhenya Huang}, \bibinfo{person}{Hao Wang},
  \bibinfo{person}{Zheyuan Hu}, \bibinfo{person}{Qi Liu},
  \bibinfo{person}{Enhong Chen}, {and} \bibinfo{person}{Cheekong Lee}.}
  \bibinfo{year}{2020}\natexlab{}.
\newblock \showarticletitle{ASGN: An active semi-supervised graph neural
  network for molecular property prediction}. In
  \bibinfo{booktitle}{\emph{\hide{Proceedings of the 26th ACM SIGKDD
  International Conference on Knowledge Discovery \& Data Mining}SIGKDD}}.
  \bibinfo{pages}{731--752}.
\newblock


\bibitem[\protect\citeauthoryear{Jacob and Vert}{Jacob and Vert}{2008}]%
        {jacob2008protein}
\bibfield{author}{\bibinfo{person}{Laurent Jacob} {and}
  \bibinfo{person}{Jean-Philippe Vert}.} \bibinfo{year}{2008}\natexlab{}.
\newblock \showarticletitle{Protein-ligand interaction prediction: an improved
  chemogenomics approach}.
\newblock \bibinfo{journal}{\emph{Bioinformatics}} \bibinfo{volume}{24},
  \bibinfo{number}{19} (\bibinfo{year}{2008}), \bibinfo{pages}{2149--2156}.
\newblock


\bibitem[\protect\citeauthoryear{Jain}{Jain}{2003}]%
        {jain2003surflex}
\bibfield{author}{\bibinfo{person}{Ajay~N Jain}.}
  \bibinfo{year}{2003}\natexlab{}.
\newblock \showarticletitle{Surflex: fully automatic flexible molecular docking
  using a molecular similarity-based search engine}.
\newblock \bibinfo{journal}{\emph{Journal of medicinal chemistry}}
  \bibinfo{volume}{46}, \bibinfo{number}{4} (\bibinfo{year}{2003}),
  \bibinfo{pages}{499--511}.
\newblock


\bibitem[\protect\citeauthoryear{Jhoti and Leach}{Jhoti and Leach}{2007}]%
        {jhoti2007structure}
\bibfield{author}{\bibinfo{person}{Harren Jhoti} {and}
  \bibinfo{person}{Andrew~R Leach}.} \bibinfo{year}{2007}\natexlab{}.
\newblock \bibinfo{booktitle}{\emph{Structure-based drug discovery}}.
  Vol.~\bibinfo{volume}{1}.
\newblock \bibinfo{publisher}{Springer}.
\newblock


\bibitem[\protect\citeauthoryear{Kinnings, Liu, Tonge, Jackson, Xie, and
  Bourne}{Kinnings et~al\mbox{.}}{2011}]%
        {kinnings2011machine}
\bibfield{author}{\bibinfo{person}{Sarah~L Kinnings}, \bibinfo{person}{Nina
  Liu}, \bibinfo{person}{Peter~J Tonge}, \bibinfo{person}{Richard~M Jackson},
  \bibinfo{person}{Lei Xie}, {and} \bibinfo{person}{Philip~E Bourne}.}
  \bibinfo{year}{2011}\natexlab{}.
\newblock \showarticletitle{A machine learning-based method to improve docking
  scoring functions and its application to drug repurposing}.
\newblock \bibinfo{journal}{\emph{Journal of chemical information and
  modeling}} \bibinfo{volume}{51}, \bibinfo{number}{2} (\bibinfo{year}{2011}),
  \bibinfo{pages}{408--419}.
\newblock


\bibitem[\protect\citeauthoryear{Kipf and Welling}{Kipf and Welling}{2017}]%
        {kipf2017semi}
\bibfield{author}{\bibinfo{person}{Thomas~N. Kipf} {and} \bibinfo{person}{Max
  Welling}.} \bibinfo{year}{2017}\natexlab{}.
\newblock \showarticletitle{Semi-Supervised Classification with Graph
  Convolutional Networks}. In \bibinfo{booktitle}{\emph{\hide{International
  Conference on Learning Representations (ICLR)}ICLR}}.
\newblock


\bibitem[\protect\citeauthoryear{Kitchen, Decornez, Furr, and Bajorath}{Kitchen
  et~al\mbox{.}}{2004}]%
        {kitchen2004docking}
\bibfield{author}{\bibinfo{person}{Douglas~B Kitchen},
  \bibinfo{person}{H{\'e}l{\`e}ne Decornez}, \bibinfo{person}{John~R Furr},
  {and} \bibinfo{person}{J{\"u}rgen Bajorath}.}
  \bibinfo{year}{2004}\natexlab{}.
\newblock \showarticletitle{Docking and scoring in virtual screening for drug
  discovery: methods and applications}.
\newblock \bibinfo{journal}{\emph{Nature reviews Drug discovery}}
  \bibinfo{volume}{3}, \bibinfo{number}{11} (\bibinfo{year}{2004}),
  \bibinfo{pages}{935--949}.
\newblock


\bibitem[\protect\citeauthoryear{Klicpera, Gro{\ss}, and
  G{\"u}nnemann}{Klicpera et~al\mbox{.}}{2020}]%
        {klicpera_dimenet_2020}
\bibfield{author}{\bibinfo{person}{Johannes Klicpera}, \bibinfo{person}{Janek
  Gro{\ss}}, {and} \bibinfo{person}{Stephan G{\"u}nnemann}.}
  \bibinfo{year}{2020}\natexlab{}.
\newblock \showarticletitle{Directional Message Passing for Molecular Graphs}.
  In \bibinfo{booktitle}{\emph{\hide{International Conference on Learning
  Representations (ICLR)}ICLR}}.
\newblock


\bibitem[\protect\citeauthoryear{Leach, Shoichet, and Peishoff}{Leach
  et~al\mbox{.}}{2006}]%
        {leach2006prediction}
\bibfield{author}{\bibinfo{person}{Andrew~R Leach}, \bibinfo{person}{Brian~K
  Shoichet}, {and} \bibinfo{person}{Catherine~E Peishoff}.}
  \bibinfo{year}{2006}\natexlab{}.
\newblock \showarticletitle{Prediction of protein- ligand interactions. Docking
  and scoring: successes and gaps}.
\newblock \bibinfo{journal}{\emph{Journal of medicinal chemistry}}
  \bibinfo{volume}{49}, \bibinfo{number}{20} (\bibinfo{year}{2006}),
  \bibinfo{pages}{5851--5855}.
\newblock


\bibitem[\protect\citeauthoryear{Leckband, Israelachvili, Schmitt, and
  Knoll}{Leckband et~al\mbox{.}}{1992}]%
        {leckband1992long}
\bibfield{author}{\bibinfo{person}{DE Leckband}, \bibinfo{person}{JN
  Israelachvili}, \bibinfo{person}{FJ Schmitt}, {and} \bibinfo{person}{W
  Knoll}.} \bibinfo{year}{1992}\natexlab{}.
\newblock \showarticletitle{Long-range attraction and molecular rearrangements
  in receptor-ligand interactions}.
\newblock \bibinfo{journal}{\emph{Science}} \bibinfo{volume}{255},
  \bibinfo{number}{5050} (\bibinfo{year}{1992}), \bibinfo{pages}{1419--1421}.
\newblock


\bibitem[\protect\citeauthoryear{Li, Leung, Wong, and Ballester}{Li
  et~al\mbox{.}}{2015}]%
        {li2015low}
\bibfield{author}{\bibinfo{person}{Hongjian Li}, \bibinfo{person}{Kwong-Sak
  Leung}, \bibinfo{person}{Man-Hon Wong}, {and} \bibinfo{person}{Pedro~J
  Ballester}.} \bibinfo{year}{2015}\natexlab{}.
\newblock \showarticletitle{Low-quality structural and interaction data
  improves binding affinity prediction via random forest}.
\newblock \bibinfo{journal}{\emph{Molecules}} \bibinfo{volume}{20},
  \bibinfo{number}{6} (\bibinfo{year}{2015}), \bibinfo{pages}{10947--10962}.
\newblock


\bibitem[\protect\citeauthoryear{Li and Shah}{Li and Shah}{2017}]%
        {PMID:28150235}
\bibfield{author}{\bibinfo{person}{Qingliang Li} {and} \bibinfo{person}{Salim
  Shah}.} \bibinfo{year}{2017}\natexlab{}.
\newblock \showarticletitle{Structure-Based Virtual Screening}.
\newblock \bibinfo{journal}{\emph{Methods in molecular biology (Clifton,
  N.J.)}}  \bibinfo{volume}{1558} (\bibinfo{year}{2017}),
  \bibinfo{pages}{111—124}.
\newblock
\showISSN{1064-3745}


\bibitem[\protect\citeauthoryear{Li, Zhou, Xu, Liu, Lu, and Xiong}{Li
  et~al\mbox{.}}{2020}]%
        {li2020competitive}
\bibfield{author}{\bibinfo{person}{Shuangli Li}, \bibinfo{person}{Jingbo Zhou},
  \bibinfo{person}{Tong Xu}, \bibinfo{person}{Hao Liu},
  \bibinfo{person}{Xinjiang Lu}, {and} \bibinfo{person}{Hui Xiong}.}
  \bibinfo{year}{2020}\natexlab{}.
\newblock \showarticletitle{Competitive Analysis for Points of Interest}. In
  \bibinfo{booktitle}{\emph{Proceedings of the 26th ACM SIGKDD International
  Conference on Knowledge Discovery \& Data Mining}}.
  \bibinfo{pages}{1265--1274}.
\newblock


\bibitem[\protect\citeauthoryear{Lim, Ryu, Park, Choe, Ham, and Kim}{Lim
  et~al\mbox{.}}{2019}]%
        {lim2019predicting}
\bibfield{author}{\bibinfo{person}{Jaechang Lim}, \bibinfo{person}{Seongok
  Ryu}, \bibinfo{person}{Kyubyong Park}, \bibinfo{person}{Yo~Joong Choe},
  \bibinfo{person}{Jiyeon Ham}, {and} \bibinfo{person}{Woo~Youn Kim}.}
  \bibinfo{year}{2019}\natexlab{}.
\newblock \showarticletitle{Predicting drug--target interaction using a novel
  graph neural network with 3D structure-embedded graph representation}.
\newblock \bibinfo{journal}{\emph{Journal of chemical information and
  modeling}} \bibinfo{volume}{59}, \bibinfo{number}{9} (\bibinfo{year}{2019}),
  \bibinfo{pages}{3981--3988}.
\newblock


\bibitem[\protect\citeauthoryear{Liu, Han, Fu, Zhou, Lu, and Xiong}{Liu
  et~al\mbox{.}}{2021}]%
        {liu2021vldb}
\bibfield{author}{\bibinfo{person}{Hao Liu}, \bibinfo{person}{Jindong Han},
  \bibinfo{person}{Yanjie Fu}, \bibinfo{person}{Jingbo Zhou},
  \bibinfo{person}{Xinjiang Lu}, {and} \bibinfo{person}{Hui Xiong}.}
  \bibinfo{year}{2021}\natexlab{}.
\newblock \showarticletitle{Multi-Modal Transportation Recommendation with
  Unified Route Representation Learning}.
\newblock \bibinfo{journal}{\emph{Proceedings of the VLDB Endowment}}
  \bibinfo{volume}{14}, \bibinfo{number}{3} (\bibinfo{year}{2021}),
  \bibinfo{pages}{342--350}.
\newblock


\bibitem[\protect\citeauthoryear{Liu, Yuan, Cai, and Ji}{Liu
  et~al\mbox{.}}{2020}]%
        {liu2020deep}
\bibfield{author}{\bibinfo{person}{Yi Liu}, \bibinfo{person}{Hao Yuan},
  \bibinfo{person}{Lei Cai}, {and} \bibinfo{person}{Shuiwang Ji}.}
  \bibinfo{year}{2020}\natexlab{}.
\newblock \showarticletitle{Deep learning of high-order interactions for
  protein interface prediction}. In \bibinfo{booktitle}{\emph{\hide{Proceedings
  of the 26th ACM SIGKDD International Conference on Knowledge Discovery \&
  Data Mining}SIGKDD}}. \bibinfo{pages}{679--687}.
\newblock


\bibitem[\protect\citeauthoryear{Maziarka, Danel, Mucha, Rataj, Tabor, and
  Jastrz{\k{e}}bski}{Maziarka et~al\mbox{.}}{2020}]%
        {maziarka2020molecule}
\bibfield{author}{\bibinfo{person}{{\L}ukasz Maziarka}, \bibinfo{person}{Tomasz
  Danel}, \bibinfo{person}{S{\l}awomir Mucha}, \bibinfo{person}{Krzysztof
  Rataj}, \bibinfo{person}{Jacek Tabor}, {and} \bibinfo{person}{Stanis{\l}aw
  Jastrz{\k{e}}bski}.} \bibinfo{year}{2020}\natexlab{}.
\newblock \showarticletitle{Molecule Attention Transformer}.
\newblock \bibinfo{journal}{\emph{arXiv preprint arXiv:2002.08264}}
  (\bibinfo{year}{2020}).
\newblock


\bibitem[\protect\citeauthoryear{Meng, Zhang, Mezei, and Cui}{Meng
  et~al\mbox{.}}{2011}]%
        {meng2011molecular}
\bibfield{author}{\bibinfo{person}{Xuan-Yu Meng}, \bibinfo{person}{Hong-Xing
  Zhang}, \bibinfo{person}{Mihaly Mezei}, {and} \bibinfo{person}{Meng Cui}.}
  \bibinfo{year}{2011}\natexlab{}.
\newblock \showarticletitle{Molecular docking: a powerful approach for
  structure-based drug discovery}.
\newblock \bibinfo{journal}{\emph{Current computer-aided drug design}}
  \bibinfo{volume}{7}, \bibinfo{number}{2} (\bibinfo{year}{2011}),
  \bibinfo{pages}{146--157}.
\newblock


\bibitem[\protect\citeauthoryear{Moitessier, Englebienne, Lee, Lawandi, and
  Corbeil}{Moitessier et~al\mbox{.}}{2008}]%
        {moitessier2008towards}
\bibfield{author}{\bibinfo{person}{N Moitessier}, \bibinfo{person}{P
  Englebienne}, \bibinfo{person}{D Lee}, \bibinfo{person}{J Lawandi}, {and}
  \bibinfo{person}{CR Corbeil}.} \bibinfo{year}{2008}\natexlab{}.
\newblock \showarticletitle{Towards the development of universal, fast and
  highly accurate docking/scoring methods: a long way to go}.
\newblock \bibinfo{journal}{\emph{British journal of pharmacology}}
  \bibinfo{volume}{153}, \bibinfo{number}{S1} (\bibinfo{year}{2008}),
  \bibinfo{pages}{S7--S26}.
\newblock


\bibitem[\protect\citeauthoryear{Muegge and Martin}{Muegge and Martin}{1999}]%
        {muegge1999general}
\bibfield{author}{\bibinfo{person}{Ingo Muegge} {and} \bibinfo{person}{Yvonne~C
  Martin}.} \bibinfo{year}{1999}\natexlab{}.
\newblock \showarticletitle{A general and fast scoring function for protein-
  ligand interactions: a simplified potential approach}.
\newblock \bibinfo{journal}{\emph{Journal of medicinal chemistry}}
  \bibinfo{volume}{42}, \bibinfo{number}{5} (\bibinfo{year}{1999}),
  \bibinfo{pages}{791--804}.
\newblock


\bibitem[\protect\citeauthoryear{Nguyen, Le, Quinn, Nguyen, Le, and
  Venkatesh}{Nguyen et~al\mbox{.}}{2020}]%
        {10.1093/bioinformatics/btaa921}
\bibfield{author}{\bibinfo{person}{Thin Nguyen}, \bibinfo{person}{Hang Le},
  \bibinfo{person}{Thomas~P Quinn}, \bibinfo{person}{Tri Nguyen},
  \bibinfo{person}{Thuc~Duy Le}, {and} \bibinfo{person}{Svetha Venkatesh}.}
  \bibinfo{year}{2020}\natexlab{}.
\newblock \showarticletitle{{GraphDTA: Predicting drug–target binding
  affinity with graph neural networks}}.
\newblock \bibinfo{journal}{\emph{Bioinformatics}} (\bibinfo{date}{10}
  \bibinfo{year}{2020}).
\newblock
\showISSN{1367-4803}
\newblock
\shownote{btaa921.}


\bibitem[\protect\citeauthoryear{{\"O}zt{\"u}rk, {\"O}zg{\"u}r, and
  Ozkirimli}{{\"O}zt{\"u}rk et~al\mbox{.}}{2018}]%
        {ozturk2018deepdta}
\bibfield{author}{\bibinfo{person}{Hakime {\"O}zt{\"u}rk},
  \bibinfo{person}{Arzucan {\"O}zg{\"u}r}, {and} \bibinfo{person}{Elif
  Ozkirimli}.} \bibinfo{year}{2018}\natexlab{}.
\newblock \showarticletitle{DeepDTA: deep drug--target binding affinity
  prediction}.
\newblock \bibinfo{journal}{\emph{Bioinformatics}} \bibinfo{volume}{34},
  \bibinfo{number}{17} (\bibinfo{year}{2018}), \bibinfo{pages}{i821--i829}.
\newblock


\bibitem[\protect\citeauthoryear{Ragoza, Hochuli, Idrobo, Sunseri, and
  Koes}{Ragoza et~al\mbox{.}}{2017}]%
        {ragoza2017protein}
\bibfield{author}{\bibinfo{person}{Matthew Ragoza}, \bibinfo{person}{Joshua
  Hochuli}, \bibinfo{person}{Elisa Idrobo}, \bibinfo{person}{Jocelyn Sunseri},
  {and} \bibinfo{person}{David~Ryan Koes}.} \bibinfo{year}{2017}\natexlab{}.
\newblock \showarticletitle{Protein--ligand scoring with convolutional neural
  networks}.
\newblock \bibinfo{journal}{\emph{Journal of chemical information and
  modeling}} \bibinfo{volume}{57}, \bibinfo{number}{4} (\bibinfo{year}{2017}),
  \bibinfo{pages}{942--957}.
\newblock


\bibitem[\protect\citeauthoryear{Song, Zheng, Niu, Fu, Lu, and Yang}{Song
  et~al\mbox{.}}{2020}]%
        {song2020communicative}
\bibfield{author}{\bibinfo{person}{Ying Song}, \bibinfo{person}{Shuangjia
  Zheng}, \bibinfo{person}{Zhangming Niu}, \bibinfo{person}{Zhang-Hua Fu},
  \bibinfo{person}{Yutong Lu}, {and} \bibinfo{person}{Yuedong Yang}.}
  \bibinfo{year}{2020}\natexlab{}.
\newblock \showarticletitle{Communicative representation learning on attributed
  molecular graphs}. In \bibinfo{booktitle}{\emph{\hide{Proceedings of the
  Twenty-Ninth International Joint Conference on Artificial Intelligence,(IJCAI
  2020)}IJCAI}}. \bibinfo{pages}{2831--2838}.
\newblock


\bibitem[\protect\citeauthoryear{Sousa, Fernandes, and Ramos}{Sousa
  et~al\mbox{.}}{2006}]%
        {sousa2006protein}
\bibfield{author}{\bibinfo{person}{Sergio~Filipe Sousa},
  \bibinfo{person}{Pedro~Alexandrino Fernandes}, {and}
  \bibinfo{person}{Maria~Joao Ramos}.} \bibinfo{year}{2006}\natexlab{}.
\newblock \showarticletitle{Protein--ligand docking: current status and future
  challenges}.
\newblock \bibinfo{journal}{\emph{Proteins: Structure, Function, and
  Bioinformatics}} \bibinfo{volume}{65}, \bibinfo{number}{1}
  (\bibinfo{year}{2006}), \bibinfo{pages}{15--26}.
\newblock


\bibitem[\protect\citeauthoryear{Stepniewska-Dziubinska, Zielenkiewicz, and
  Siedlecki}{Stepniewska-Dziubinska et~al\mbox{.}}{2018}]%
        {stepniewska2018development}
\bibfield{author}{\bibinfo{person}{Marta~M Stepniewska-Dziubinska},
  \bibinfo{person}{Piotr Zielenkiewicz}, {and} \bibinfo{person}{Pawel
  Siedlecki}.} \bibinfo{year}{2018}\natexlab{}.
\newblock \showarticletitle{Development and evaluation of a deep learning model
  for protein--ligand binding affinity prediction}.
\newblock \bibinfo{journal}{\emph{Bioinformatics}} \bibinfo{volume}{34},
  \bibinfo{number}{21} (\bibinfo{year}{2018}), \bibinfo{pages}{3666--3674}.
\newblock


\bibitem[\protect\citeauthoryear{Su, Yang, Du, Feng, Liu, Li, and Wang}{Su
  et~al\mbox{.}}{2018}]%
        {su2018comparative}
\bibfield{author}{\bibinfo{person}{Minyi Su}, \bibinfo{person}{Qifan Yang},
  \bibinfo{person}{Yu Du}, \bibinfo{person}{Guoqin Feng},
  \bibinfo{person}{Zhihai Liu}, \bibinfo{person}{Yan Li}, {and}
  \bibinfo{person}{Renxiao Wang}.} \bibinfo{year}{2018}\natexlab{}.
\newblock \showarticletitle{Comparative assessment of scoring functions: the
  CASF-2016 update}.
\newblock \bibinfo{journal}{\emph{Journal of chemical information and
  modeling}} \bibinfo{volume}{59}, \bibinfo{number}{2} (\bibinfo{year}{2018}),
  \bibinfo{pages}{895--913}.
\newblock


\bibitem[\protect\citeauthoryear{Sun, Zhao, Gilvary, Elemento, Zhou, and
  Wang}{Sun et~al\mbox{.}}{2020}]%
        {sun2020graph}
\bibfield{author}{\bibinfo{person}{Mengying Sun}, \bibinfo{person}{Sendong
  Zhao}, \bibinfo{person}{Coryandar Gilvary}, \bibinfo{person}{Olivier
  Elemento}, \bibinfo{person}{Jiayu Zhou}, {and} \bibinfo{person}{Fei Wang}.}
  \bibinfo{year}{2020}\natexlab{}.
\newblock \showarticletitle{Graph convolutional networks for computational drug
  development and discovery}.
\newblock \bibinfo{journal}{\emph{Briefings in bioinformatics}}
  \bibinfo{volume}{21}, \bibinfo{number}{3} (\bibinfo{year}{2020}),
  \bibinfo{pages}{919--935}.
\newblock


\bibitem[\protect\citeauthoryear{Trott and Olson}{Trott and Olson}{2010}]%
        {trott2010autodock}
\bibfield{author}{\bibinfo{person}{Oleg Trott} {and} \bibinfo{person}{Arthur~J
  Olson}.} \bibinfo{year}{2010}\natexlab{}.
\newblock \showarticletitle{AutoDock Vina: improving the speed and accuracy of
  docking with a new scoring function, efficient optimization, and
  multithreading}.
\newblock \bibinfo{journal}{\emph{Journal of computational chemistry}}
  \bibinfo{volume}{31}, \bibinfo{number}{2} (\bibinfo{year}{2010}),
  \bibinfo{pages}{455--461}.
\newblock


\bibitem[\protect\citeauthoryear{Veli{\v{c}}kovi{\'c}, Cucurull, Casanova,
  Romero, Li{\`o}, and Bengio}{Veli{\v{c}}kovi{\'c} et~al\mbox{.}}{2018}]%
        {velivckovic2018graph}
\bibfield{author}{\bibinfo{person}{Petar Veli{\v{c}}kovi{\'c}},
  \bibinfo{person}{Guillem Cucurull}, \bibinfo{person}{Arantxa Casanova},
  \bibinfo{person}{Adriana Romero}, \bibinfo{person}{Pietro Li{\`o}}, {and}
  \bibinfo{person}{Yoshua Bengio}.} \bibinfo{year}{2018}\natexlab{}.
\newblock \showarticletitle{Graph Attention Networks}. In
  \bibinfo{booktitle}{\emph{\hide{International Conference on Learning
  Representations}ICLR}}.
\newblock


\bibitem[\protect\citeauthoryear{Wallach, Dzamba, and Heifets}{Wallach
  et~al\mbox{.}}{2015}]%
        {wallach2015atomnet}
\bibfield{author}{\bibinfo{person}{Izhar Wallach}, \bibinfo{person}{Michael
  Dzamba}, {and} \bibinfo{person}{Abraham Heifets}.}
  \bibinfo{year}{2015}\natexlab{}.
\newblock \showarticletitle{AtomNet: a deep convolutional neural network for
  bioactivity prediction in structure-based drug discovery}.
\newblock \bibinfo{journal}{\emph{arXiv preprint arXiv:1510.02855}}
  (\bibinfo{year}{2015}).
\newblock


\bibitem[\protect\citeauthoryear{Wang, Fang, Lu, Yang, and Wang}{Wang
  et~al\mbox{.}}{2005}]%
        {wang2005pdbbind}
\bibfield{author}{\bibinfo{person}{Renxiao Wang}, \bibinfo{person}{Xueliang
  Fang}, \bibinfo{person}{Yipin Lu}, \bibinfo{person}{Chao-Yie Yang}, {and}
  \bibinfo{person}{Shaomeng Wang}.} \bibinfo{year}{2005}\natexlab{}.
\newblock \showarticletitle{The PDBbind database: methodologies and updates}.
\newblock \bibinfo{journal}{\emph{Journal of medicinal chemistry}}
  \bibinfo{volume}{48}, \bibinfo{number}{12} (\bibinfo{year}{2005}),
  \bibinfo{pages}{4111--4119}.
\newblock


\bibitem[\protect\citeauthoryear{Wang, Lai, and Wang}{Wang
  et~al\mbox{.}}{2002}]%
        {wang2002further}
\bibfield{author}{\bibinfo{person}{Renxiao Wang}, \bibinfo{person}{Luhua Lai},
  {and} \bibinfo{person}{Shaomeng Wang}.} \bibinfo{year}{2002}\natexlab{}.
\newblock \showarticletitle{Further development and validation of empirical
  scoring functions for structure-based binding affinity prediction}.
\newblock \bibinfo{journal}{\emph{Journal of computer-aided molecular design}}
  \bibinfo{volume}{16}, \bibinfo{number}{1} (\bibinfo{year}{2002}),
  \bibinfo{pages}{11--26}.
\newblock


\bibitem[\protect\citeauthoryear{Xu, Hu, Leskovec, and Jegelka}{Xu
  et~al\mbox{.}}{2019}]%
        {xu2018powerful}
\bibfield{author}{\bibinfo{person}{Keyulu Xu}, \bibinfo{person}{Weihua Hu},
  \bibinfo{person}{Jure Leskovec}, {and} \bibinfo{person}{Stefanie Jegelka}.}
  \bibinfo{year}{2019}\natexlab{}.
\newblock \showarticletitle{How Powerful are Graph Neural Networks?}. In
  \bibinfo{booktitle}{\emph{\hide{International Conference on Learning
  Representations}ICLR}}.
\newblock


\bibitem[\protect\citeauthoryear{Yang, Swanson, Jin, Coley, Eiden, Gao,
  Guzman-Perez, Hopper, Kelley, Mathea, et~al\mbox{.}}{Yang
  et~al\mbox{.}}{2019}]%
        {yang2019analyzing}
\bibfield{author}{\bibinfo{person}{Kevin Yang}, \bibinfo{person}{Kyle Swanson},
  \bibinfo{person}{Wengong Jin}, \bibinfo{person}{Connor Coley},
  \bibinfo{person}{Philipp Eiden}, \bibinfo{person}{Hua Gao},
  \bibinfo{person}{Angel Guzman-Perez}, \bibinfo{person}{Timothy Hopper},
  \bibinfo{person}{Brian Kelley}, \bibinfo{person}{Miriam Mathea},
  {et~al\mbox{.}}} \bibinfo{year}{2019}\natexlab{}.
\newblock \showarticletitle{Analyzing learned molecular representations for
  property prediction}.
\newblock \bibinfo{journal}{\emph{Journal of chemical information and
  modeling}} \bibinfo{volume}{59}, \bibinfo{number}{8} (\bibinfo{year}{2019}),
  \bibinfo{pages}{3370--3388}.
\newblock


\bibitem[\protect\citeauthoryear{Zang and Wang}{Zang and Wang}{2020}]%
        {zang2020moflow}
\bibfield{author}{\bibinfo{person}{Chengxi Zang} {and} \bibinfo{person}{Fei
  Wang}.} \bibinfo{year}{2020}\natexlab{}.
\newblock \showarticletitle{MoFlow: an invertible flow model for generating
  molecular graphs}. In \bibinfo{booktitle}{\emph{\hide{Proceedings of the 26th
  ACM SIGKDD International Conference on Knowledge Discovery \& Data
  Mining}SIGKDD}}. \bibinfo{pages}{617--626}.
\newblock


\bibitem[\protect\citeauthoryear{Zheng, Fan, and Mu}{Zheng
  et~al\mbox{.}}{2019}]%
        {zheng2019onionnet}
\bibfield{author}{\bibinfo{person}{Liangzhen Zheng}, \bibinfo{person}{Jingrong
  Fan}, {and} \bibinfo{person}{Yuguang Mu}.} \bibinfo{year}{2019}\natexlab{}.
\newblock \showarticletitle{OnionNet: a Multiple-Layer
  Intermolecular-Contact-Based Convolutional Neural Network for Protein--Ligand
  Binding Affinity Prediction}.
\newblock \bibinfo{journal}{\emph{ACS omega}} \bibinfo{volume}{4},
  \bibinfo{number}{14} (\bibinfo{year}{2019}), \bibinfo{pages}{15956--15965}.
\newblock


\bibitem[\protect\citeauthoryear{Zheng, Wang, Xu, Shen, Qin, Huai, Liu, and
  Chen}{Zheng et~al\mbox{.}}{2021}]%
        {zheng2021drug}
\bibfield{author}{\bibinfo{person}{Zhi Zheng}, \bibinfo{person}{Chao Wang},
  \bibinfo{person}{Tong Xu}, \bibinfo{person}{Dazhong Shen},
  \bibinfo{person}{Penggang Qin}, \bibinfo{person}{Baoxing Huai},
  \bibinfo{person}{Tongzhu Liu}, {and} \bibinfo{person}{Enhong Chen}.}
  \bibinfo{year}{2021}\natexlab{}.
\newblock \showarticletitle{Drug Package Recommendation via Interaction-aware
  Graph Induction}.
\newblock \bibinfo{journal}{\emph{arXiv preprint arXiv:2102.03577}}
  (\bibinfo{year}{2021}).
\newblock


\bibitem[\protect\citeauthoryear{Zhou, Li, Huang, Xiong, Wang, Xu, Xiong, and
  Dou}{Zhou et~al\mbox{.}}{2020}]%
        {zhou2020distance}
\bibfield{author}{\bibinfo{person}{Jingbo Zhou}, \bibinfo{person}{Shuangli Li},
  \bibinfo{person}{Liang Huang}, \bibinfo{person}{Haoyi Xiong},
  \bibinfo{person}{Fan Wang}, \bibinfo{person}{Tong Xu}, \bibinfo{person}{Hui
  Xiong}, {and} \bibinfo{person}{Dejing Dou}.} \bibinfo{year}{2020}\natexlab{}.
\newblock \showarticletitle{Distance-aware Molecule Graph Attention Network for
  Drug-Target Binding Affinity Prediction}.
\newblock \bibinfo{journal}{\emph{arXiv preprint arXiv:2012.09624}}
  (\bibinfo{year}{2020}).
\newblock


\end{thebibliography}

\appendix
\section{Appendix}
In the appendix, we first introduce the construction process of complex interaction graph. Then the details of experimental settings and baseline descriptions are given.
Finally, we show the additional results of parameter analysis on another dataset. 
The pseudocode of \model training procedure is described in Algorithm \ref{alg-training}. 
The implement our model based on PaddlePaddle\footnote{https://github.com/PaddlePaddle/Paddle}.  We train all models on 24 Intel CPUs and a Tesla V100 GPU with 32 GB memory.
The code is available at: \url{https://github.com/PaddlePaddle/PaddleHelix/tree/dev/apps/drug_target_interaction/sign}.

\subsection{Complex Interaction Graph Construction}
\label{a-graph-constrcut}
As shown in Figure \ref{fig-graph}(a), \old{there is no \hide{natural bond}available intermolecular connection information between the ligand and its protein in the dataset. What's more, the local covalent bonds in the original molecular graph cannot provide adequate 3D structure information.} To include non-local correlations, here we aim to construct the spatial-based complex interaction graph. Since the size of a protein is much larger than that of a ligand, it's unnecessary to build the complete protein structure in the complex graph, which might be noisy and time-consuming. Therefore, we apply a sampling-based process to construct the graph, which preserves the key structure of complex. The detail of the construction process is provided in algorithm \ref{alg-graph}. We first initialize the atom node set $\mathcal{V}$ with ligand's atom set \ligandV. Then the protein's atoms which are close to the ligand from \proteinV are selected to add into the atom node set $\mathcal{V}$. Finally, we update the complex edge set $\mathcal{E}$ by adding into the edges of \hide{complex }atom pairs whose distances are smaller than the cutoff threshold $r_{\theta}$.

\vspace{-2mm}
\normalem
\begin{algorithm}
    \SetAlgoLined
    \SetKwInOut{Input}{Input}
    \SetKwInOut{Output}{Output}
    \Input{The position matrix \proteinM\  and node set \proteinV of protein \newline 
    The position matrix \ligandM\  and node set \ligandV\  of ligand \newline
    The cutoff distance $r_{\theta}$}
    \Output{The complex interaction graph $\graph=<\mathcal{V},\mathcal{E}>$}
    Initialize $\mathcal{V} \leftarrow \ligandV$, $\mathcal{E} \leftarrow \{\}$\;
    \For{atom node pair $(a_i,\ a_j) \in \ligandV \times \proteinV$}{
        Calculate distance $d_{ij} \leftarrow |\ligandM(a_i) - \proteinM(a_j)|$\;
        \If{$d_{ij} \le r_{\theta}$}{
            Update node set $\mathcal{V} \leftarrow \mathcal{V} \cup \{a_j\}$;
        }
    }
    Combined position matrix $M \leftarrow CONCAT(\ligandM, \proteinM)$\;
    \For{atom node pair $(a_i,\ a_j) \in \mathcal{V} \times \mathcal{V}$}{
        Calculate distance $d_{ij} \leftarrow |M(a_i) - M(a_j)|$\;
        \If{$d_{ij} \le r_{\theta}$}{
            Update edge set $\mathcal{E} \leftarrow \mathcal{E} \cup \{e_{ij}=(a_i,a_j)\}$;
        }
    }
    \textbf{return} $\mathcal{V},\mathcal{E}$
    \caption{Complex Interaction Graph Construction.}
    \label{alg-graph}
\end{algorithm}
\vspace{-3mm}

\subsection{Instruction of the Binding Affinity}
\label{a-pka}
In biology experiments, the binding affinity between protein and ligand can be determined as the value $K_d$ (dissociation constant), $K_i$ (inhibition constant), or $IC_{50}$ (half inhibition concentration). \hide{Generally}In practice, the experimental binding affinity (i.e., the ground truth for the binding affinity prediction task) is expressed with the negative logarithm $pk_a$ of the determined value (e.g, $-logK_d$, $-logK_i$, or $-logIC_{50}$) on PDBbind and CSAR-HiQ datasets.

\subsection{Experiment Details}
\label{a-implement}

\subsubsection{Evaluation Metrics}
We first detail the four metrics used in our experiment. Root Mean Square Error (RMSE), Mean Absolute Error (MAE) and Pearson correlation coefficient (R) are defined as:
\begin{equation}
    RMSE = \sqrt{\frac{1}{|\mathcal{D}|}\sum_{i=1}^{|\mathcal{D}|} (\hat{y}_i-y_i)^2},\ MAE = \frac{1}{|\mathcal{D}|}\sum_{i=1}^{|\mathcal{D}|} |\hat{y}_i-y_i| 
\end{equation}
\begin{equation}
    R = \frac{\sum_{i=1}^{|\mathcal{D}|}(\hat{y}_i-\bar{\hat{y}})(y_i-\bar{y})}{\sqrt{\sum_{i=1}^{|\mathcal{D}|}(\hat{y}_i-\bar{\hat{y}})^2(y_i-\bar{y})^2}}
\end{equation}
$\hat{y}_i$ and $y_i$ respectively represent the predicted and experimental \hide{value}binding affinity of the $i$-th complex in dataset $\mathcal{D}$. As introduced in \cite{stepniewska2018development}, the standard deviation (SD) is defined as follows:
\begin{equation}
    SD = \sqrt{\frac{1}{|\mathcal{D}|-1}\sum_{i=1}^{|\mathcal{D}|} [y_i - (a+b \hat{y}_i)]^2}
\end{equation}
where $a$ and $b$ are the intercept and the slope of the regression line, respectively. 


\normalem
\begin{algorithm}[t]
    \SetAlgoLined
    \SetKwInOut{Input}{Input}
    \SetKwInOut{Output}{Output}
    \Input{Training set $\mathcal{D}$}
    \Output{Trained model parameters $\theta$}
    Randomly initialize the parameter $\theta$ of \model;
    
    \For{iteration = 1, 2, ...}{
        \For{each batch from training samples}{
            Calculate spatial relation embeddings $\bm{d}$ using Eq. (\ref{eq-embed})\;
            \For{l = 1...L}{
                Obtain edge embeddings $\bm{e}^{(l)}$ using Eq. (\ref{eq-a2e})-(\ref{eq-e-combine})\;
                Obtain node embeddings $\bm{a}^{(l)}$ using Eq. (\ref{eq-node-trans})-(\ref{eq-e2a})\;
            }
            Build the \hide{global}interaction matrix $\bm{Z}$ using Eq. (\ref{eq-im1})-(\ref{eq-im2})\;
            Estimate the interaction matrix $\tilde{\bm{Z}}$ using Eq. (\ref{eq-pool})-(\ref{eq-im-normal})\;
            Calculate the interaction loss $\mathcal{L}_z$ using Eq. (\ref{eq-inter-loss})\;
            Calculate the prediction $\hat{y}$ using Eq. (\ref{eq-predict})\;
            Calculate the prediction loss $\mathcal{L}_a$ using Eq. (\ref{eq-pred-loss})\;
            Update parameters $\theta$ according to the gradient of $\mathcal{L}$\;
        }
    }
    \textbf{return} $\theta$
    \caption{Training Procedure for \model.}
    \label{alg-training}
\end{algorithm}
\subsubsection{Input Graph and Features}
For all GNN-based methods, we use the same input complex graph as introduced in Appendix \ref{a-graph-constrcut} for protein-ligand binding affinity prediction. For GraphDTA models, we input the protein sequence as well as the ligand molecular graph or our constructed complex graph. In this paper, we report the best result when using the complex graph as input. \old{For 3D-CNN and GNN models, the atom features used according to \cite{stepniewska2018development} include atomic types, hybridization, the number of bonds with other heavy-atoms and hetero-atoms, atom properties such as aromaticity, and the partial charge. In total, 18 features are used to describe an atom. Considering the heterogeneity in the complex graph, we further extend atom features to a 36-dimension vector with zero-padding, where the 1st to 18th elements represent the features of ligand atoms and the 19th to 36th elements represent the features of protein atoms.} As to edge features for edge-based GNN models, we combine the two atom features and encoded distance features between atoms as the input vector. In brief, we provide the same input complex graph and atomic features for our \model and all GNN-based baselines to make fair comparisons. For ML-based methods and OnionNet, they cannot take the complex graph or grid-like data as input and only receive the specific molecule-level features. We extract the input feature vectors based on the distance as described in their original papers \cite{ballester2010machine, zheng2019onionnet}. Note that these extracted features also reflect the structural information of complex in a global view.

\begin{figure}
\setlength{\abovecaptionskip}{2.mm}
\setlength{\belowcaptionskip}{-0.cm}
  \centering
  \subfigure{
    \label{parameter-core-layer} 
    \includegraphics[width=0.47\columnwidth]{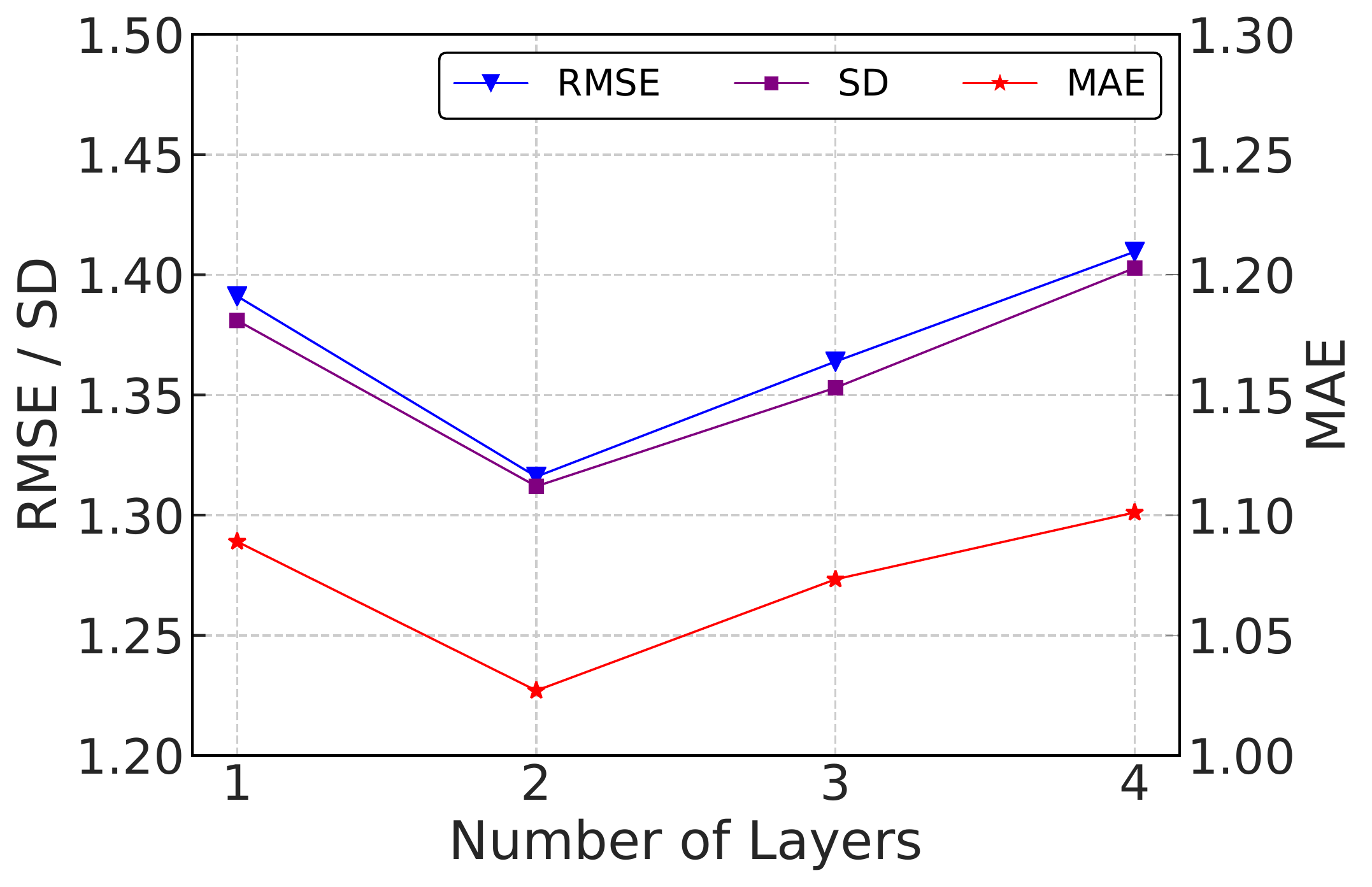}}
      \subfigure{
    \label{parameter-core-lambda} 
    \includegraphics[width=0.47\columnwidth]{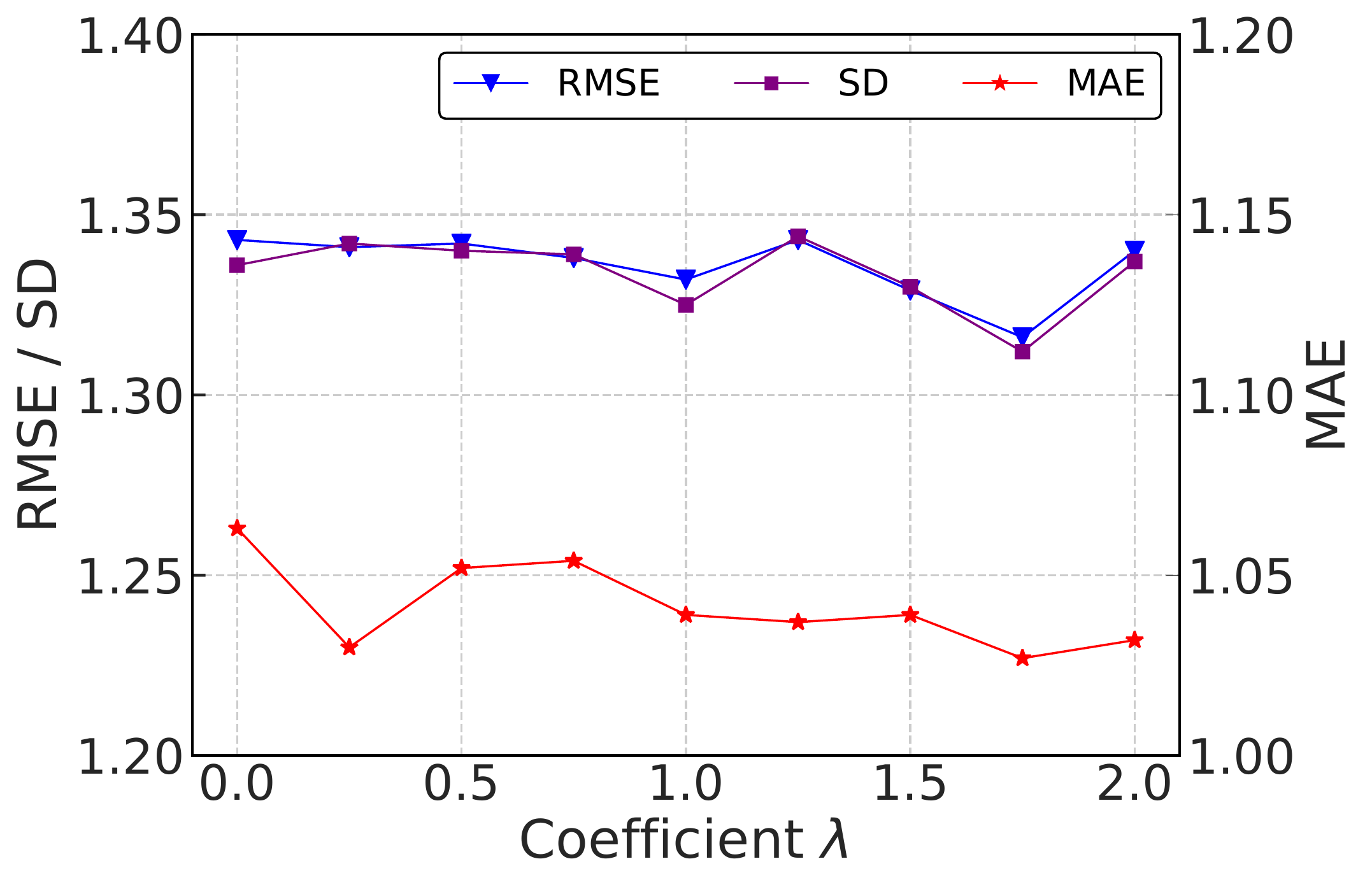}}
  \vspace{-2mm}
  \caption{Parameter analysis on PDBbind \textit{core set}.}
  \vspace{-4mm}
  \label{fig-parameter-add} 
\end{figure}


\subsubsection{Parameter Settings}
For the proposed \model, we use Adam optimizer for model training with a learning rate of 0.001 and set the batch size as 32. The balancing hyper-parameter $\lambda$ is set to 1.75 according to the performance on validation set. We construct the complex graph and interaction matrix with cutoff-threshold $\theta_d=5Å$ and $\theta_{\rho}=12Å$ as suggested in \cite{muegge1999general}, respectively. The basic dimensions of node and edge embeddings are both set to 128. The number of buckets for spatial relation $b$ is set to 4 with the splitting granularity of 1Å. For \gnn layers, we set the number of attention heads $C$ to 4, the dropout rate to 0.2, and the number of angle domains $N$ to 6. For \pool layer, there are 36 pooling blocks in total, where the two atomic type sets $S_P$ and $S_L$ are defined as stated in \cite{ballester2010machine}. 

For baseline models, we tune the parameters based on the default settings to get optimal performance. Specifically, the number of decision trees in RF-score is set to 100, the max-depth of trees is set to 5, the maximum number of features is set to 3 and the minimum number of samples required to split is set to 10. For the CNN-based models, we set the channels of three-layer 3D convolutions for Pafnucy as 64,128 and 256. For OnionNet, the number of input features is 3840 and there are 32, 64, and 128 filters in the three convolutional layers with the kernel size as 4. The maximum length of protein sequences is set to 1000 in GraphDTA. For GNN-based models, the number of filters in SGCN is set to 32 with the dimension as 36. We also apply the data augmentation process to ensure optimal performance. For fair comparison, the embedding dimension of other baselines is set to 128 (same as \model). For GNN-DTI, the initial $\mu$ and $\delta$ for distance learning in GAT layers are set to 4.0 and 1.0, respectively. For DimeNet, the number of spherical harmonics and radial basis functions are set to 4 and 3, respectively. We use two-layer interaction blocks and three-layer bilinear layers to make DimeNet work in our experiment. For DMPNN and CMPNN, the layer of edge-oriented message passing layers is set to 3 and we use MLP as the communication module in CMPNN. The weighting 
coefficients for self-attention, distance, and adjacency matrices in MAT are set to 0.3, 0.3, and 0.4, respectively.
\vspace{-2mm}
\subsection{Baseline Descriptions}
We compare our \model model with the following methods to predict the protein-ligand binding affinity:
\label{a-baseline}
\begin{itemize}[leftmargin=*,topsep=3pt]
    \item \B{ML-based methods} include linear regression (LR), support vector regression (SVR), and random forest (RF). These methods take the inter-molecular interaction features introduced in RF-Score \cite{ballester2010machine} as input and predict the protein-ligand binding affinity.
    \item \B{Pafnucy} \cite{stepniewska2018development} is a representative 3D CNN-based model which can learn the spatial structure of protein-ligand complexes.
    \item \B{OnionNet} \cite{zheng2019onionnet} generates two-dimensional interaction features based on rotation-free element-pair contacts in complexes and adopts CNN to learn representations for prediction.
    \item \B{GraphDTA} \cite{10.1093/bioinformatics/btaa921} introduces GNN models to learn the complex graph and uses CNN to learn the protein sequence. It has four variants with different GNN models: \B{GCN} \cite{kipf2017semi}, \B{GAT} \cite{velivckovic2018graph}, \B{GIN} \cite{xu2018powerful} and \B{GAT-GCN} which combines the former two models.
    \item \B{SGCN} \cite{danel2020spatial} leverages node positions based on graph convolutional network, which directly utilizes atomic coordinates.
    \item \B{GNN-DTI} \cite{lim2019predicting} is a distance-aware graph attention network with considering 3D structural information to learn the intermolecular interactions for protein-ligand complexes.
    \item \B{DMPNN} \cite{yang2019analyzing} is an edge-based message passing neural network. It can incorporate the spatial information between atoms by applying the aggregation process for edges.
    \item \B{MAT} \cite{maziarka2020molecule} employs a molecule-augmented attention mechanism based on transformer for graph representation learning with using the inter-atomic distances.
    \item \B{DimeNet} \cite{klicpera_dimenet_2020} is a recent state-of-the-art model for small molecular graph learning using directional message passing scheme. Bessel functions are employed to encode the angle and distance information in graph neural network. 
    \item \B{CMPNN} \cite{song2020communicative} further develops DMPNN to build a communicative message passing scheme between nodes and edges for better molecular representation learning.
\end{itemize}
\vspace{-2mm}
\subsection{Additional Parameters Analysis}
\label{a-para}


\B{Number of \gnn layers $L$}. As shown in Figure \ref{fig-parameter-add}, we first present the influence of multi-hop propagation with stacking node-edge interaction layers from 1 to 4. We observe that increasing the number of layers would not always give rise to a better result. The model with one \gnn layer has limited ability to model high-order information in the complex. As a result of over-fitting, the performance of the model using more than 3 layers starts to degenerate gradually. Therefore, applying two interaction layers in \model is enough to capture sufficient spatial information.

\B{Balancing coefficient $\lambda$}. Moreover, we change the coefficient $\lambda$ to control the trade-off between the prediction loss and interaction loss. From the results, we observe that the performance first tends to get better with incorporating more interactive information for long-range dependencies, and then begins to drop off slightly. In general, our model is stable with varying coefficients and always achieves better performance than all baseline methods.

\end{document}